\documentclass[sigconf,natbib=false,dvipsnames,printacmref=false]{acmart}
\RequirePackage{filecontents}
\RequirePackage[acm]{comgipp}
\RequirePackage{amsfonts}
\RequirePackage{amsmath}

\AtBeginDocument{%
  \providecommand\BibTeX{{%
    \normalfont B\kern-0.5em{\scshape i\kern-0.25em b}\kern-0.8em\TeX}}}   

\setcopyright{iw3c2w3}
\copyrightyear{2020}
\acmYear{2020}
\acmConference[WWW\,'20]{Proceedings of The Web Conference 2020}{April 20--24, 2020}{Taipei, Taiwan}
\acmBooktitle{Proceedings of The Web Conference 2020 (WWW\,'20), April 20--24, 2020, Taipei, Taiwan}
\acmPrice{}
\acmDOI{10.1145/3366423.3380218}
\acmISBN{978-1-4503-7023-3/20/04}
\acmSubmissionID{fp1321}

\usepackage[T1]{fontenc}
\usepackage{float}

\usepackage[
	backend=biber,
	style=numeric,
	url=false,
	isbn=false,
	maxbibnames=99, %
 	sorting=ynt
	]{biblatex}
\AtEveryBibitem{%
	\clearname{translator}%
	\clearfield{pagetotal}%
	\clearfield{publisher}%
	\clearfield{issn}%
	\clearfield{groups}%
	\clearfield{volume}
	\clearfield{number}
	\clearname{editor}
}

\DeclareFieldFormat
[article,inbook,incollection,inproceedings,patent,thesis,unpublished]
{title}{#1}

\ifdefined\theReferenceText
\fi

  \newcommand{\theReferenceText}{\fullcite{GreinerPetter2020}}

\usepackage{graphicx} 
\graphicspath{img/} %
\usepackage{caption}
\usepackage{subcaption}
\usepackage{wrapfig}
\usepackage{array}
\usepackage{multirow}
\usepackage{booktabs} %

\usepackage{lstomdoc}
\usepackage{relsize}
\newcommand\AVGsub[1]{\operatorname{AVG_{\mbox{\textscale{0.4}{\,#1}}}}}

\newcommand{\Jacobi}[4]{P_{#1}^{(#2, #3)}\!\left(#4\right)}
\newcommand{\StandardJacobi}{\Jacobi{n}{\alpha}{\beta}{x}}
\providecommand{\MathML}{\textsc{MathML}}%
\makeatletter
\newcommand{\LaTeXML}{{L\kern-.36em{\sbox\z@ T\vbox to\ht\z@{\hbox{\check@mathfonts\fontsize\sf@size\z@\math@fontsfalse\selectfont A}\vss}}\kern-.15em T\kern-.1667em\lower.4ex\hbox{E}\kern-0.05em\relax{\scshape xml}}}
\makeatother

\newcommand\TB{\rule[-1.5ex]{0pt}{0pt}}
\newcommand\TT{\rule{0pt}{3.2ex}}

\begin{document}

\title{Discovering Mathematical Objects of Interest---A Study of Mathematical Notations}

\renewcommand{\shorttitle}{Discovering Mathematical Objects of Interest}
\newcommand{\aff}[1]{\texorpdfstring{$^{#1}$}{}}
\author{Andr\'{e} Greiner-Petter\aff{1},  Moritz Schubotz\aff{1,2},  Fabian  M\"{u}ller\aff{2},  
Corinna Breitinger\aff{1,5}, Howard S.~Cohl\aff{3}, Akiko Aizawa\aff{4}, Bela Gipp\aff{1,5}}
\affiliation{%
  $^1$ University of Wuppertal, Germany ({andre.greiner-petter@zbmath.org, \{last\}@uni-wuppertal.de})\\
  $^2$ FIZ-Karlsruhe, Germany (\{first.last\}@fiz-karlsruhe.de)\\
  $^3$ National Institute of Standards and Technology, U.S.A (\{first.last\}@nist.gov)\\
  $^4$ National Institute of Informatics, Japan (\{last\}@nii.ac.jp)\\
  $^5$ University of Konstanz, Germany (\{first.last\}@uni-konstanz.de)
}

\renewcommand{\shortauthors}{A.~Greiner-Petter, et al.}

\begin{abstract}
	Mathematical notation, i.e., the writing system used to communicate concepts in mathematics, encodes valuable information for a variety of information search and retrieval systems. Yet, mathematical notations remain mostly unutilized by today's systems. In this paper, we present the first in-depth study on the distributions of mathematical notation in two large scientific corpora:~the open access arXiv (2.5B mathematical objects) and the mathematical reviewing service for
	pure and applied mathematics zbMATH (61M mathematical objects). Our study lays a foundation for future research projects on mathematical information retrieval for large scientific corpora. Further, we demonstrate the relevance of our results to a variety of use-cases. For example, to assist semantic extraction systems, to improve scientific search engines, and to facilitate specialized math recommendation systems.
	
	The contributions of our presented research are as follows:
	(1) we present 
	the first distributional analysis of mathematical formulae on arXiv and zbMATH;
	(2) we retrieve relevant mathematical objects for given textual search queries (e.g., linking $P_{n}^{(\alpha, \beta)}\!\left(x\right)$ with `Jacobi polynomial');
	(3) we extend zbMATH's search engine by providing relevant mathematical formulae; and
	(4) we exemplify the applicability of the results by presenting auto-completion for math inputs as the first contribution to math recommendation systems.
	To expedite future research projects, we have made available our source code and data.
\end{abstract}


\maketitle
\thispagestyle{firststyle}

\vspace{-0.45cm}
\section{Introduction}\label{sec:intro}
Taking into account mathematical notation in the literature leads to a better understanding of scientific literature on the Web and allows one to make use of semantic information in specialized Information Retrieval (IR) systems. 
Nowadays applications in Math Information Retrieval (MathIR)~\cite{GuidiC16}, such as search engines~\cite{LohiaSVK05,kamaliT10,Kohlhase2012,KamaliT13,kristiantoTHA14,OhashiKTA16,DavilaZ17}, semantic extraction systems~\cite{Schubotz16,Kristianto2017,Schubotz2017}, recent efforts in math embeddings~\cite{Gao17,Blei2018,Greiner-PetterR19,Youssef19}, and semantic tagging of math formulae~\cite{ChienC15,POM-Tagger} either consider an entire equation as one entity or only focus on single symbols.
Since math expressions often contain meaningful and important subexpressions, these applications could benefit from an approach that lies between the extremes of examining only individual symbols or considering an entire equation as one entity.
Consider for example, the explicit definition for Jacobi polynomials
\cite[(18.5.7)]{NIST:DLMF}
\setlength{\abovedisplayskip}{0pt}
\setlength{\belowdisplayskip}{0pt}
\begin{equation}\label{eq:jacobi-def}
\medmuskip=0mu
\thinmuskip=0mu
\thickmuskip=0mu
\hspace{-0.1cm}P_n^{(\alpha,\beta)}(x) = \frac{\Gamma(\alpha+n+1)}{n!\,\Gamma(\alpha+\beta+n+1)} \sum_{m=0}^{n} \binom{n}{m} \frac{\Gamma(\alpha+\beta+n+m+1)}{\Gamma(\alpha+m+1)} \left( \frac{x-1}{2} \right)^m.
\end{equation}
The \textit{interesting} components in this equation are $P_n^{(\alpha,\beta)}(x)$ on the 
left-hand side, and the appearance of the gamma function $\Gamma(s)$ on the 
right-hand side, implying a direct relationship between Jacobi polynomials and the gamma function.
Considering the entire expression as a single object misses this important relationship.
On the other hand, focusing on single symbols can result in the misleading interpretation of $\Gamma$ as a variable and $\Gamma(\alpha + n + 1)$ as a multiplication between $\Gamma$ and $(\alpha + n + 1)$.
A system capable of identifying the important components, such as $P_n^{(\alpha,\beta)}(x)$ or $\Gamma(\alpha + n + 1)$, is therefore desirable. Hereafter, we define these components as \emph{Mathematical Objects of Interest} (MOIs)~\cite{Greiner-PetterR19}.

The \textit{importance} of math objects is a somewhat imprecise description and thus difficult to measure. 
Currently, not much effort has been made in identifying meaningful subexpressions.
Kristianto et al.~\cite{Kristianto2017} introduced dependency graphs between formulae.
With this approach, they were able to build dependency graphs of mathematical expressions, but only if the expressions appeared as single expressions in the context.
For example, if $\Gamma(\alpha + n + 1)$ appears as a stand-alone expression in the context, the algorithm will declare a dependency with Equation~\eqref{eq:jacobi-def}.
However, it is more likely that different forms, such as $\Gamma(s)$, appear in the context.
Since this expression does not match any subexpression in Equation~\eqref{eq:jacobi-def}, the approach cannot establish a connection with $\Gamma(s)$.
Kohlhase et al.~studied in~\cite{AKohlhase2017,AKohlhase18a,AKohlhaseKO18} another approach to identify essential components in formulae. 
They performed eye-tracking studies to identify important areas in rendered mathematical formulae.
While this is an interesting approach that allows one to learn more about the insights of human behaviors of reading and understanding math, it is inaccessible for extensive studies. 

This paper presents the first extensive frequency distribution study of mathematical equations in two large scientific corpora, the e-Print archive arXiv.org (hereafter referred to as arXiv\footnote{\url{https://arxiv.org/} 
[Accessed:~Sep.~1, 2019]}) and the international reviewing service for pure and applied mathematics
zbMATH\footnote{\url{https://zbmath.org} [Accessed:~Sep.~1, 2019]}. 
We will show that math expressions, similar to words in natural language corpora, also obey Zipf's law~\cite{Piantadosi2014}, and therefore follows a \emph{Zipfian} distribution. 
Related research projects observed a relation to Zipf's law for single math symbols~\cite{ChienC15,Schubotz16}.
In the context of quantitative linguistics, Zipf's law states that given a text corpus, the frequency of any word is inversely proportional to its rank in the frequency table. 
Motivated by the similarity to linguistic properties, we will present a novel approach for ranking formulae by their relevance via a customized version of the ranking function BM25~\cite{RobertsonZ09}.
We will present results that can be easily embedded in other systems in order to distinguish between 
common and uncommon notations within formulae.
Our results lay a foundation for future research projects in MathIR.

Fundamental knowledge on frequency distributions of math formulae is beneficial for numerous applications in MathIR, ranging from educational purposes~\cite{Smith04} to math recommendation systems, search engines~\cite{OhashiKTA16,DavilaZ17}, and even automatic plagiarism detection systems~\cite{MeuschkeSHSG17,Meuschke2019,Schubotz2019}.
For example, students can search for the conventions to write certain quantities in formulae; document preparation systems can integrate an auto-completion or auto-correction service for math inputs; search or recommendation engines can adjust their ranking scores with respect to standard notations; and plagiarism detection systems can estimate whether two identical formulae indicate potential plagiarism or are just using the conventional notations in a particular subject area.
To exemplify the applicability of our findings, we present a textual search approach to retrieve mathematical formulae.
Further, we will extend zbMATH's faceted search by providing facets of mathematical formulae according to a given textual search query. 
Lastly, we present a simple auto-completion system for math inputs as a contribution towards advancing mathematical recommendation systems. 
Further, we show that the results provide useful insights for plagiarism detection algorithms.
We provide access to the source code, the results, and extended versions of all of the figures appearing in this paper at \url{https://github.com/ag-gipp/FormulaCloudData}.

\noindent
\textit{Related Work:}
Today, mathematical search engines index formulae in a database. Much effort has been undertaken to make this process as efficient as possible in terms of precision and runtime performance~\cite{LohiaSVK05,kamaliT10,LipaniAPLH14,Zanibbi2016,DavilaZ17}.
The generated databases naturally contain the information required to examine the distributions of the indexed mathematical formulae. 
Yet, no in-depth studies of these distributions have been undertaken. 
Instead, math search engines focus on other aspects, such as devising novel similarity measures and improving runtime efficiency. This is because the goal of math search engines is to retrieve relevant (i.e., similar) formulae which correspond to a given search query that partially~\cite{kristiantoTHA14,LipaniAPLH14,OhashiKTA16} or exclusively~\cite{kamaliT10,KamaliT13,DavilaZ17} contains formulae. 
However, for a fundamental study of distributions of mathematical expressions, no similarity measures nor efficient lookup or indexing is required. Thus, we use the general-purpose query language XQuery and employ the BaseX\footnote{\url{http://basex.org/} [Accessed:~Sep.~2019]; We used BaseX 9.2 for our experiments.} implementation.
BaseX is a free open-source XML database engine, which is fully compatible with the latest XQuery standard~\cite{BaseX1,BaseX2}. Since our implementations rely on XQuery, we are able to switch to any other database which 
allows for processing via XQuery.

\vspace{-.5cm}
\section{Data Preparation}\label{sec:data-preparation}
\LaTeX{} is the de facto standard for the preparation of academic manuscripts in the fields of mathematics and physics~\cite{Gaudeul07}.
Since \LaTeX{} allows for advanced customizations and even computations, it is challenging to process.
For this reason, \LaTeX{} expressions are unsuitable for an extensive distribution analysis of mathematical notations.
For mathematical expressions on the web, the XML formatted \MathML{}\footnote{\url{https://www.w3.org/TR/MathML3/} [Accessed:~Sep.~1, 2019]} is the current standard, as specified by the World Wide Web Consortium (W3C).
The tree structure and the fixed standard, i.e., \MathML{} tags, cannot be changed, thus making this data format reliable. %
Several available tools are able to convert from \LaTeX{} to \MathML{}~\cite{Schubotz2018c}
and various databases are able to index XML data.
Thus, for this study, we have chosen to focus on \MathML{}. 
In the following, we investigate the databases arXMLiv (08/2018)~\cite{SML:arXMLiv:08.2018} and zbMATH\footnote{\url{https://zbmath.org/} [Accessed:~Sep.~1, 2019]}~\cite{zbMATH}. 

The arXMLiv dataset ($\approx$1.2 million documents) contains HTML5 versions of the documents from the e-Print archive arXiv.org. 
The HTML5 documents were generated from the \TeX{} sources via \LaTeXML~\cite{LaTeXML}. 
\LaTeXML{} converted all mathematical expressions into \MathML{} with parallel markup, i.e., presentation and content \MathML{}. 
In this study we only consider the subsets \textit{no-problem} and \textit{warning}, which generated no errors during the conversion process. 
Nonetheless, the \MathML{} data generated still contains some errors or falsely annotated math. 
For example, we discovered several instances of affiliation and footnotes, SVG\footnote{Scalable Vector Graphics} 
and other unknown tags, %
encoded in \MathML{}. 
Regarding the footnotes, we presumed that authors falsely used mathematical environments for generating footnote or affiliation marks.
We used the \TeX{} string, provided as an attribute in the \MathML{} data, to filter out expressions that match the string `\verb|{}^{*}|', where `\verb|*|' indicates any possible expression.
In addition, we filtered out SVG and other unknown tags.
We assume that these expressions were generated by mistake due to limitations of \LaTeXML. 
The final arXiv dataset consisted of 841,008 documents which contained at least one mathematical formula. The dataset contained a total of 294,151,288 mathematical expressions. 

In addition to arXiv, we investigated zbMATH, an international reviewing service for pure and applied mathematics which contains abstracts and reviews of articles, hereafter uniformly called abstracts, mainly from the domains of pure and applied mathematics. 
The abstracts in zbMATH are formatted in \TeX{}~\cite{zbMATH}. 
To be able to compare arXiv and zbMATH, we manually generated \MathML{} via \LaTeXML{} for each mathematical formula in zbMATH and performed the same filters as used for the arXiv documents. 
The zbMATH dataset contained 2,813,451 abstracts, of which 1,349,297 contained at least one formula. In total, the dataset contained 11,747,860 formulae. 
Even though the total number of formulae is smaller compared to arXiv, we hypothesize that math formulae in abstracts are particularly meaningful.

\vspace*{-0.57cm}
\subsection{Data Wrangling}\label{sec:data-wrangling}
\setlength{\columnsep}{6pt}%
\begin{wrapfigure}{r}{0.145\textwidth}
  \vspace*{-1.7cm}
  \linespread{0.85}
\begin{lstlisting}[label={lst.MathML},numbersep=4pt,xleftmargin=15pt,mathescape=true,escapeinside={(*}{*)},caption=MathML representation of $\StandardJacobi$.]
<math><mrow>
 <msubsup>
  <mi>P</mi>
  <mi>n</mi>
  <mrow>
   <mo>(</mo>
   <mi>$\alpha$</mi>
   <mo>,</mo>
   <mi>$\beta$</mi>
   <mo>)</mo>
   <mo></mo>
  </mrow>
 </msubsup>
 <mo></mo>
 <mrow>
  <mo>(</mo>
  <mi>x</mi>
  <mo>)</mo>
 </mrow>
</mrow></math>
\end{lstlisting}
  \vspace*{-.03cm}
\end{wrapfigure}

Since we focused on the frequency distributions of visual expressions, we only considered presentational \MathML{} (pMML).
Rather than normalizing the pMML data, e.g., via MathMLCan~\cite{MathMLCan}, which would also change the tree structure and visual core elements in pMML, we only eliminated the attributes.
These attributes are used for minor visual changes, e.g., stretched parentheses or inline limits of sums and integrals.
Thus, for this first study, we preserved the core structure of the pMML data, which might 
provide insightful statistics for the \MathML{} community to further cultivate the standard.
After extracting all \MathML{} expressions, filtering out falsely annotated math and SVG tags, and eliminating unnecessary attributes and annotations, the datasets required 83GB of disk space for arXiv and 6GB for zbMATH, respectively.

In the following, we indexed the data via BaseX. 
The indexed datasets required a disk space of 143.9GB in total (140GB for arXiv and 3.9GB for zbMATH). 
Due to the limitations\footnote{A detailed overview of the limitations of BaseX databases can be found at \url{http://docs.basex.org/wiki/Statistics} [Accessed:~Sep.~1, 2019].} of databases in BaseX, it was necessary to split our datasets into smaller subsets.
We split the datasets according to the 20 major article categories of arXiv\footnote{The arXiv categories \textit{astro-ph} (astro physics), \textit{cond-mat} (condensed matter), and \textit{math} (mathematics) were still too large for a single database. Thus, we split those categories into two equally sized parts.} and classifications of zbMATH. 
To increase performance, we use BaseX in a server-client environment. 
We experienced performance issues in BaseX when multiple clients repeatedly requested data from the same server in short intervals. 
We determined that the best workaround for this issue was to launch BaseX servers for each database, i.e., each category/classification. 

Mathematical expressions often consist of multiple meaningful subexpressions, which we defined as MOIs.
However, without further investigation of the context, it is impossible to determine meaningful subexpressions. As a consequence, every equation is a potential MOI on its own and potentially consists of multiple other MOIs.
For an extensive frequency distributional analysis, we aim to discover all possible mathematical objects. Hence, we split every formula into its components. Since \MathML{} is an XML data format (essentially a tree-structured 
format), we define subexpressions of equations as subtrees of its \MathML{} format.

Listing~\ref{lst.MathML} illustrates a Jacobi polynomial $\StandardJacobi$ in pMML. 
The \verb|<mo>| element on line 14 contains the \textit{invisible times} UTF-8 character. 
By definition, the \verb|<math>| element is the root element of \MathML{} expressions. 
Since we cut off all other elements besides pMML nodes, each \verb|<math>| element has one and only one child element\footnote{Sequences are always nested in an \texttt{<mrow>} element.}.
Thus, we define the child element of the \verb|<math>| element as the root of the expression. 
Starting from this root element, we explore all subexpressions.
For this study, we presume that every meaningful mathematical object (i.e., MOI) must contain at least one identifier.

Hence, we only study subtrees which contain at least one \verb|<mi>| node. 
Identifiers, in the sense of \MathML{}, are 
`\!\textit{symbolic names or arbitrary text}'
\footnote{\url{https://www.w3.org/TR/MathML3/chapter3.html} [Accessed:~Sep.~1, 2019]}, e.g., single Latin or Greek letters.
Identifiers do not contain special characters (other than Greek letters) or numbers. 
As a consequence, arithmetic expressions, such as $(1+2)^2$, or sequences of special characters and numbers, such as $\{1,2,...\} \cap \{-1\}$, will not appear in our distributional analysis. 
However, if a sequence or arithmetic expression consists of an identifier somewhere in the pMML tree (such as in $\{1,2,...\} \cap A$), the entire expression will be recognized.
The Jacobi polynomial $\StandardJacobi$, therefore consists of the following subexpressions:~$P_n^{(\alpha,\beta)}$, $(\alpha,\beta)$, $(x)$, and the single identifiers $P$, $n$, $\alpha$, $\beta$, and $x$.
The entire expression is also a mathematical object.
Hence, we take entire expressions with an identifier into account for our analysis. 
In the following, the set of subexpressions will be understood to include the expression itself.

For our experiments, we also generated a string representation of the \MathML{} data.
The string is generated recursively by applying one of two rules for each node:
(i) if the current node is a leaf, the node-tag and the content will be merged by a colon, e.g., \verb|<mi>x</mi>| will be converted to \verb|mi:x|; 
(ii) otherwise the node-tag wraps parentheses around its content and separates the children by a comma, e.g., \verb|<mrow><mo>(</mo><mi>x</mi><mo>)</mo></mrow>| will be converted to \verb|mrow(mo:(,mi:x,mo:))|.
Furthermore, the special UTF-8 characters for invisible times (U+2062) and function application (U+2061) are replaced by \verb|ivt| and \verb|fa|, respectively.
For example, the gamma function with argument $x+1$, $\Gamma(x+1)$ would be represented by
\begin{equation}\small
\verb|mrow(mi:|\Gamma\verb|,mo:ivt,mrow(mo:(,mrow(mi:x,mo:+,mn:1),mo:)))|.
\end{equation}
Between $\Gamma$ and $(x+1)$, there would most likely be the special character for \textit{invisible times} rather than for \textit{function application}, because \LaTeXML{} is not able to parse $\Gamma$ as a function.
Note that this string conversion is a bijective mapping.
The string representation reduces the verbose XML format to a more concise presentation. Thus, an equivalence check between two expressions is more efficient.

\vspace{-.3cm}
\subsection{Complexity of Math}
Mathematical expressions can become complex and lengthy.
The tree structure of \MathML{} allows us to introduce a measure that reflects the complexity of mathematical expressions.
More complex expressions usually consist of more extensively nested subtrees in the \MathML{} data.
Thus, we define the complexity of a mathematical expression by the maximum depth of the \MathML{} tree. 
In XML  the content of a node and its attributes are commonly interpreted as children of the node. 
Thus, we define the depth of a single node as 1 rather than 0, i.e., single identifiers, such as \verb|<mi>P</mi>|, have a complexity of 1.
The Jacobi polynomial from Listing~\ref{lst.MathML} has a complexity of 4.

We perform the extraction of subexpressions from \MathML{} in BaseX.
The algorithm for the extraction process is written in XQuery.
The algorithm traverses recursively downwards from the root to the leaves. 
In each iteration, it checks whether there is an identifier, i.e., \verb|<mi>| element, among the descendants 
of the current node. 
If there is no such element, the subtree will be ignored. 
It seems counterintuitive to start from the root and check if an identifier is among the descendants rather than starting at each identifier and traversing upwards to the root.
If an XQuery requests a node in BaseX, BaseX loads the entire subtree of the requested node into the 
cache (up to a specified size). 
If the algorithm traverses upwards through the \MathML{} tree, the XQuery will trigger database requests in every iteration. 
Hence, the downwards implementation performs better, since there is only one database request for every expression rather than for every subexpression.

Since we only minimize the pMML data rather than normalizing it, two identically rendered expressions may have different complexities.
For instance, \verb|<mrow><mi>x</mi></mrow>| consists of two distinct subexpressions,
but both of them are displayed the same.
Another problem often appears for arrays or similar visually complicated structures.
The extracted expressions are not necessarily logical subexpressions.
We will consider applying more advanced embedding techniques such as special tokenizers~\cite{LipaniAPLH14}, symbol layout trees~\cite{Zanibbi2016,DavilaZ17}, and a \MathML{} normalization via MathMLCan~\cite{MathMLCan} in future research to overcome these issues.

\section{Frequency Distributions of Mathematical Formulae}
By splitting each formula into subexpressions, we generated longer documents and a bias towards low complexities. 
Note that, hereafter, we only refer to the mathematical content of documents. 
Thus, the length of a document refers to the number of math formulae---here the number 
of subexpressions---in the document.
After splitting expressions into subexpressions, arXiv consists of $2.5$B and zbMATH of $61$M expressions, which raised the average document length to $2,\!982.87$ for arXiv and $45.47$ for zbMATH, respectively. 

\begin{table}
	\vspace*{-0.3cm}
	\centering
	\begin{tabular}{lrr}
	\\\hline
	Category & arXiv & zbMATH \\\hline
	Documents & 841,008 & 1,349,297 \\
	Formulae & 294,151,288 & 11,747,860 \\
	Subexpressions & 2,508,620,512 & 61,355,307 \\
	Unique Subexpressions & 350,206,974 & 8,450,496 \\
	Average Document Length & 2,982.87 & 45.47 \\
	Average Complexity & 5.01 & 3.89 \\
	Maximum Complexity & 218 & 26 \\\hline
	\end{tabular}
	\caption{Dataset overview. Average Document Length is defined as the average number of subexpressions per document.}
	\label{tab:summary}
	\vspace*{-0.9cm}
\end{table} 

\begin{figure}[tbp]
	\centering
  	\includegraphics[width=0.45\textwidth]{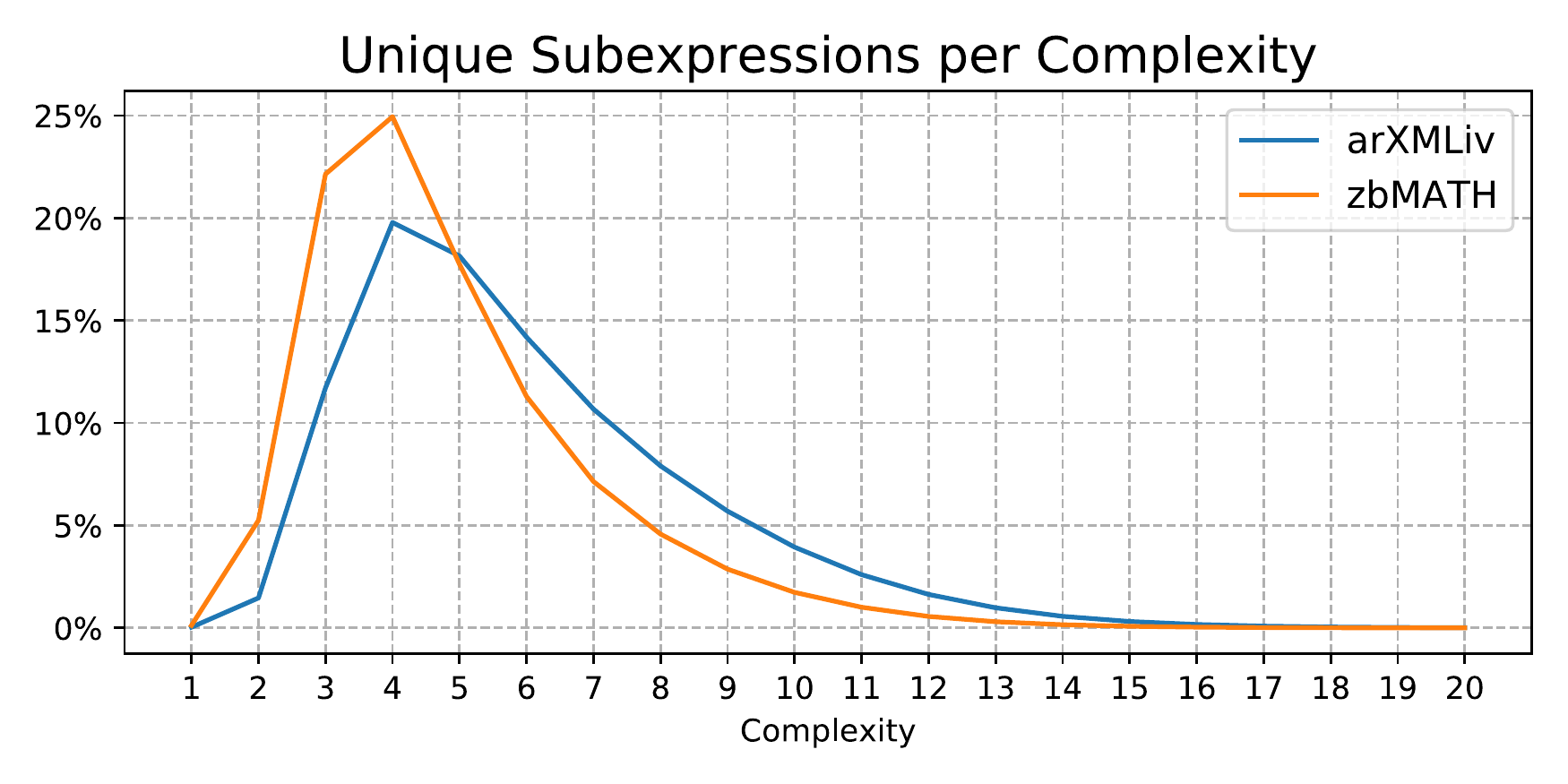}
  	\vspace*{-0.6cm}
  	\caption{Unique subexpressions for each complexity in arXiv and zbMATH.}
  	\label{fig:numPerComplexity}
  	\vspace*{-0.6cm}
\end{figure}

For calculating frequency distributions, we merged two subexpressions if their string representations were identical. Remember, the string representation is unique for each \MathML{} tree. After merging, arXiv consisted of 350,206,974 unique mathematical subexpressions with a maximum complexity of 218 and an average complexity of $5.01$. For high complexities over 70, the formulae show some erroneous structures that might be generated from \LaTeXML{} by mistake. For example, the expression with the highest complexity is a long sequence of a polynomial starting with `$P_4(t_1,t_3,t_7,t_{11}) = $' followed by 690 summands. The complexity is caused by a high number of unnecessarily deeply nested \verb|<mrow>| nodes. The highest complexity with a minimum document frequency of two is 39, which is a continued fraction. Since continued fractions are nested fractions, they naturally have a large complexity. One of the most complex expressions (complexity 20) with a minimum document frequency of three was the formula
\vspace*{-0.29cm}
\setlength{\abovedisplayskip}{0pt}
\setlength{\belowdisplayskip}{-3pt}
\begin{equation}\label{eq:hardy-littlewood}\small
\medmuskip=0mu
\thinmuskip=0mu
\thickmuskip=0mu
\left( \sum_{j_1=1}^{n} \left( \sum_{j_2=1}^{n} \left( \cdots
\left( \sum_{j_m=1}^{n} \left| T \left( e_{j_1}, \ldots, e_{j_m} \right) \right|^{q_m} \right)^{\scriptstyle\frac{q_{m-1}}{q_m}}
\cdots \right)^{\scriptstyle\frac{q_2}{q_3}} \right)^{\scriptstyle\frac{q_1}{q_2}} \right)^{\scriptstyle\frac{1}{q_1}}\mkern-15mu \leq C_{m,p,\text{\textbf{q}}}^{\mathbb{K}} \left \| T \right \|.
\vspace*{-0.15cm}
\end{equation}
In contrast, zbMATH only consisted of 8,450,496 unique expressions with a maximum complexity of 26 and an average complexity of $3.89$. 
One of the most complex expressions in zbMATH with a minimum document frequency of three was
\setlength{\abovedisplayskip}{0pt}
\setlength{\belowdisplayskip}{3pt}
\begin{equation}
M_p(r,f) = \left( \frac{1}{2\pi} \int_0^{2\pi} \left| f \left(re^{i\theta} \right) \right|^p d\theta \right)^{1/p}.
\end{equation}
As we expected, reviews and abstracts in zbMATH were generally shorter and consisted of less complex mathematical formulae. The dataset also appeared to contain fewer erroneous expressions, since expressions of complexity 25 are still readable and meaningful.

Figure~\ref{fig:numPerComplexity} shows the ratio of unique subexpressions for each complexity in both datasets.
The figure illustrates that both datasets share a peak at complexity four. 
Compared to zbMATH, the arXiv expressions are slightly more evenly distributed over the different levels of complexities.
Interestingly, complexities one and two are not dominant in either of the two datasets.
Single identifiers only make up $0.03\%$ in arXiv and $0.12\%$ in zbMATH, which is comparable to expressions of complexity 19 and 14, respectively.
This finding illustrates the problem of capturing semantic meanings for single identifiers rather than for more complex expressions~\cite{Schubotz2017}.
It also substantiates that entire expressions, if too complex, are not suitable either for capturing the semantic meanings~\cite{Kristianto2017}.
Instead, a middle ground is desirable, since the most unique expressions in both datasets have a complexity between 3 and 5.
Table~\ref{tab:summary} summarizes the statistics of the examined datasets.

\begin{figure}[tbp]
\vspace*{-0.3cm}
	\begin{subfigure}[b]{.236\textwidth}
		\centering
  		\includegraphics[width=\textwidth]{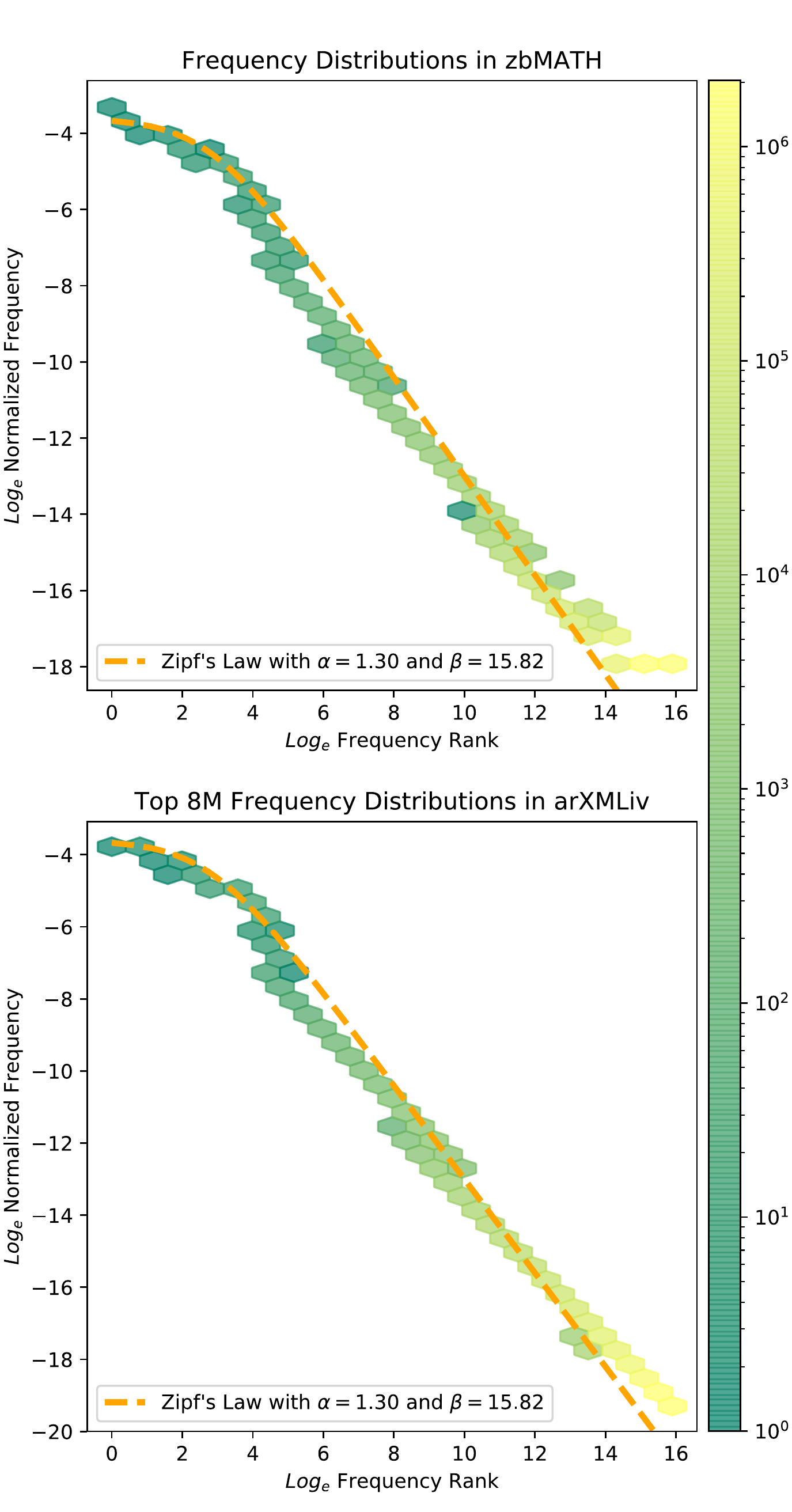}
  		\vspace*{-0.6cm}
  		\caption{Frequency Distributions}
  		\label{fig:freq}
	\end{subfigure}
	\begin{subfigure}[b]{.236\textwidth}
		\centering
  		\includegraphics[width=\textwidth]{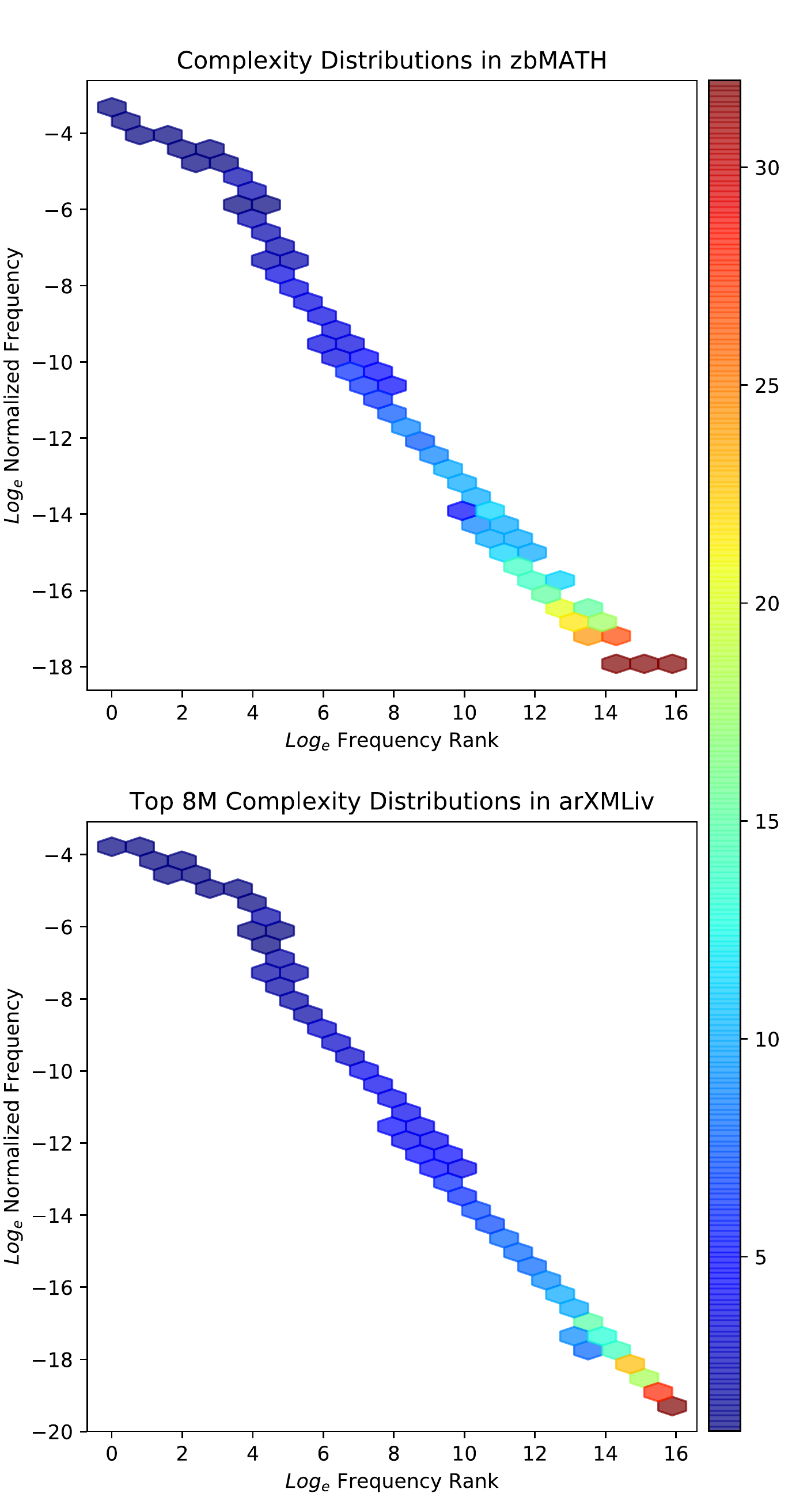}
  		\vspace*{-0.6cm}
  		\caption{Complexity Distributions}
  		\label{fig:comp}
	\end{subfigure}
	\vspace*{-0.6cm}
	\caption{Each figure illustrates the relationship between the frequency ranks ($x$-axis) and the normalized frequency ($y$-axis) in zbMATH (top) and arXiv (bottom). For arXiv, only the first 8 million entries are plotted to be comparable with zbMATH ($\approx$\,8.5 million entries). Subfigure~(\subref{fig:freq}) shades the hexagonal bins from green to yellow using a logarithmic scale according to the number of math expressions that fall into a bin. The dashed orange line represents Zipf's distribution~\eqref{eq:zipf-2}. The values for $\alpha$ and $\beta$ are provided in the plots. Subfigure~(\subref{fig:comp}) shades the bins from blue to red according to the maximum complexity in each bin.}
	\label{fig:zipf}
	\vspace*{-0.6cm}
\end{figure}

\subsection{Zipf's Law}
In linguistics, it is well known that word distributions follow Zipf's Law~\cite{Piantadosi2014}, i.e., the $r$-th most frequent word has a frequency that scales to
\vspace*{-0.15cm}
\begin{equation}\label{eq:zipf-1}
f(r) \propto \frac{1}{r^\alpha}
\end{equation}
with $\alpha \approx1$.
A better approximation can be applied by a shifted distribution\vspace*{-0.15cm}
\begin{equation}\label{eq:zipf-2}
f(r) \propto \frac{1}{(r+\beta)^\alpha},
\end{equation}
where $\alpha \approx1$ and $\beta \approx2.7$.
In a study on Zipf's law,
Piantadosi~\cite{Piantadosi2014} illustrated that not only 
words in natural language corpora
follow this law surprisingly accurately, but also many other human-created sets.
For instance, in programming languages, in biological systems, and even in music.
Since mathematical communication has derived as the result of centuries of research, it would not be surprising 
if mathematical notations would also follow Zipf's law. 
The primary conclusion of the law illustrates that there are some very common tokens against a large number of symbols which are not used frequently. 
Based on this assumption, we can postulate that a score based on frequencies might be able to measure the peculiarity of a token. 
The infamous TF-IDF ranking functions and their derivatives~\cite{Aizawa03,RobertsonZ09} have performed well in linguistics for many years and are still widely used in retrieval systems~\cite{BeelGLB16}.
However, since we split every expression into its subexpressions, we generated an anomalous bias towards shorter, i.e., less complex, formulae.
Hence, distributions of subexpressions may not obey Zipf's law.

Figure~\ref{fig:zipf} visualizes a comparison between Zipf's law and the frequency distributions of mathematical subexpressions in arXiv and zbMATH.
The dashed orange line visualizes the power law~\eqref{eq:zipf-2}.
The plots demonstrate that the distributions in both datasets obey this power law.
Interestingly, there is not much difference in the distributions between both datasets.
Both distributions seem to follow the same power law, with $\alpha = 1.3$ and $\beta=15.82$.
Moreover, we can observe that the developed complexity measure seems to be appropriate, since the complexity distributions for formulae are similar to the distributions for the length of words~\cite{Piantadosi2014}.
In other words, more complex formulae, as well as long words in natural languages, are generally more specialized and thus appear less frequent throughout the corpus.
Note that colors of the bins for complexities fluctuate for rare expressions because the color represents the maximum rather than the average complexity in each bin.
\begin{figure*}[htp]
	\centering
	\vspace*{-0.2cm}
	\begin{subfigure}[b]{.196\textwidth}
		\centering
  		\includegraphics[width=\textwidth]{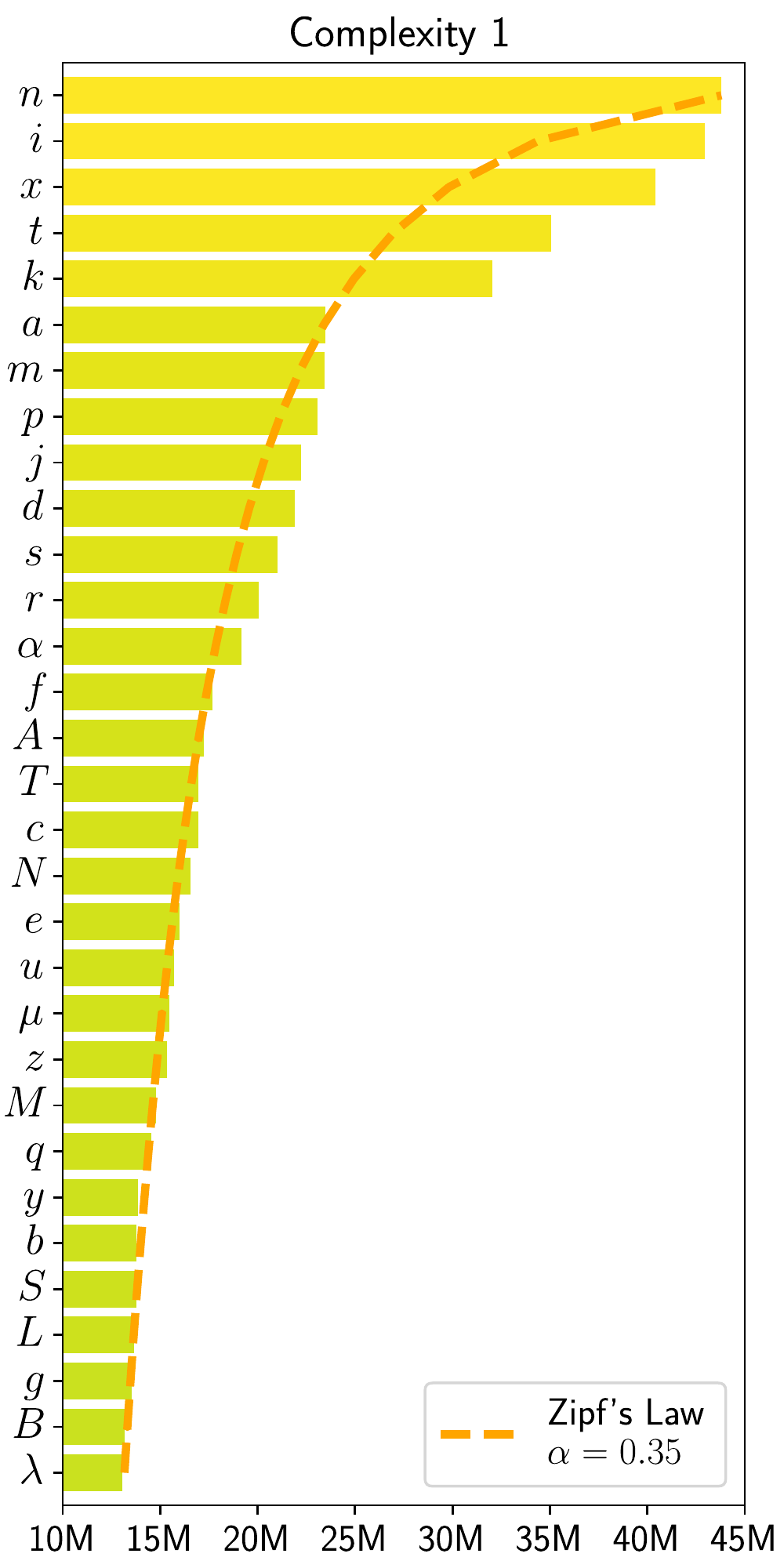}
	\end{subfigure}
  	\begin{subfigure}[b]{.196\textwidth}
		\centering
  		\includegraphics[width=\textwidth]{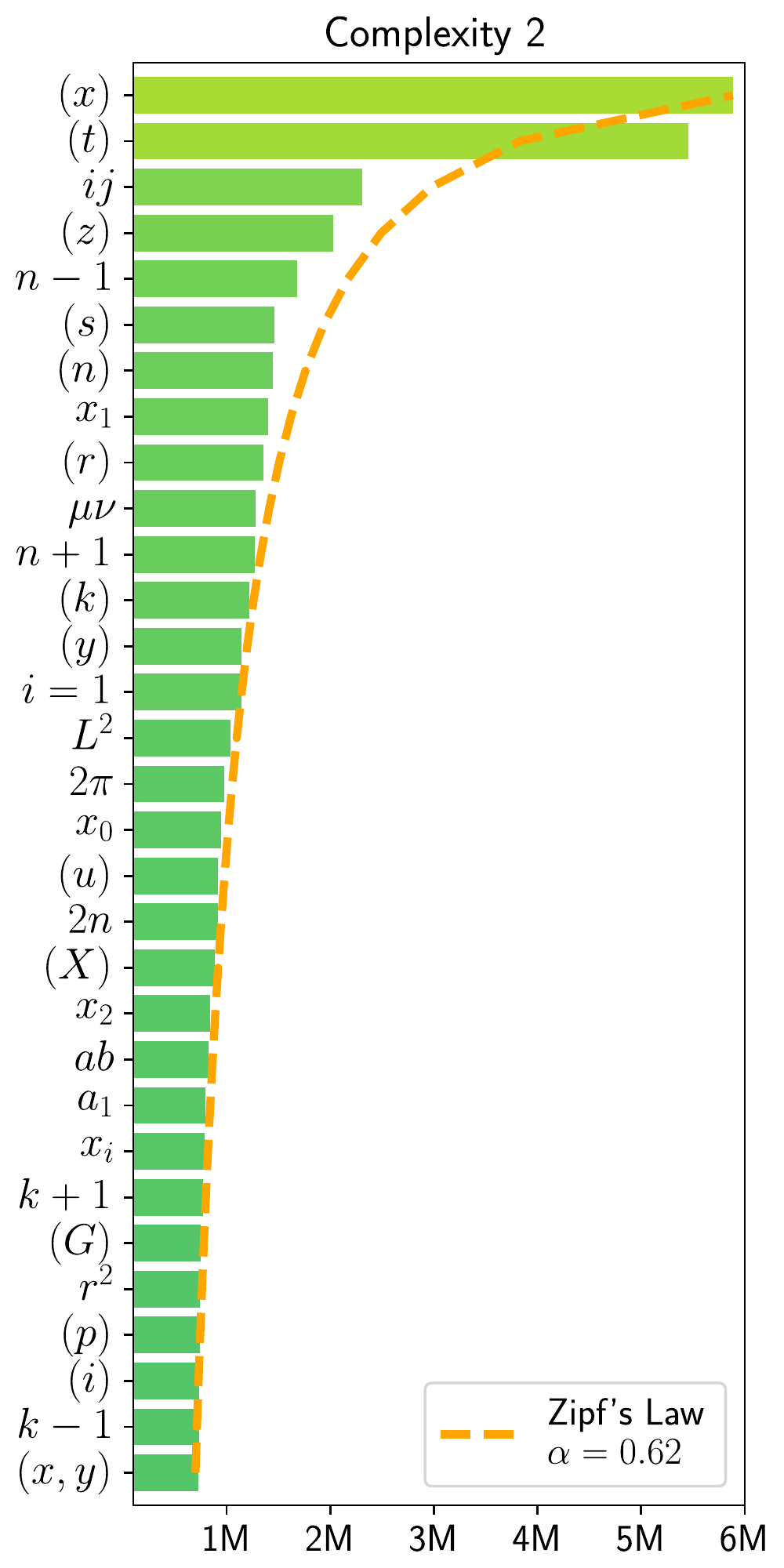}
	\end{subfigure}
  	\begin{subfigure}[b]{.196\textwidth}
		\centering
  		\includegraphics[width=\textwidth]{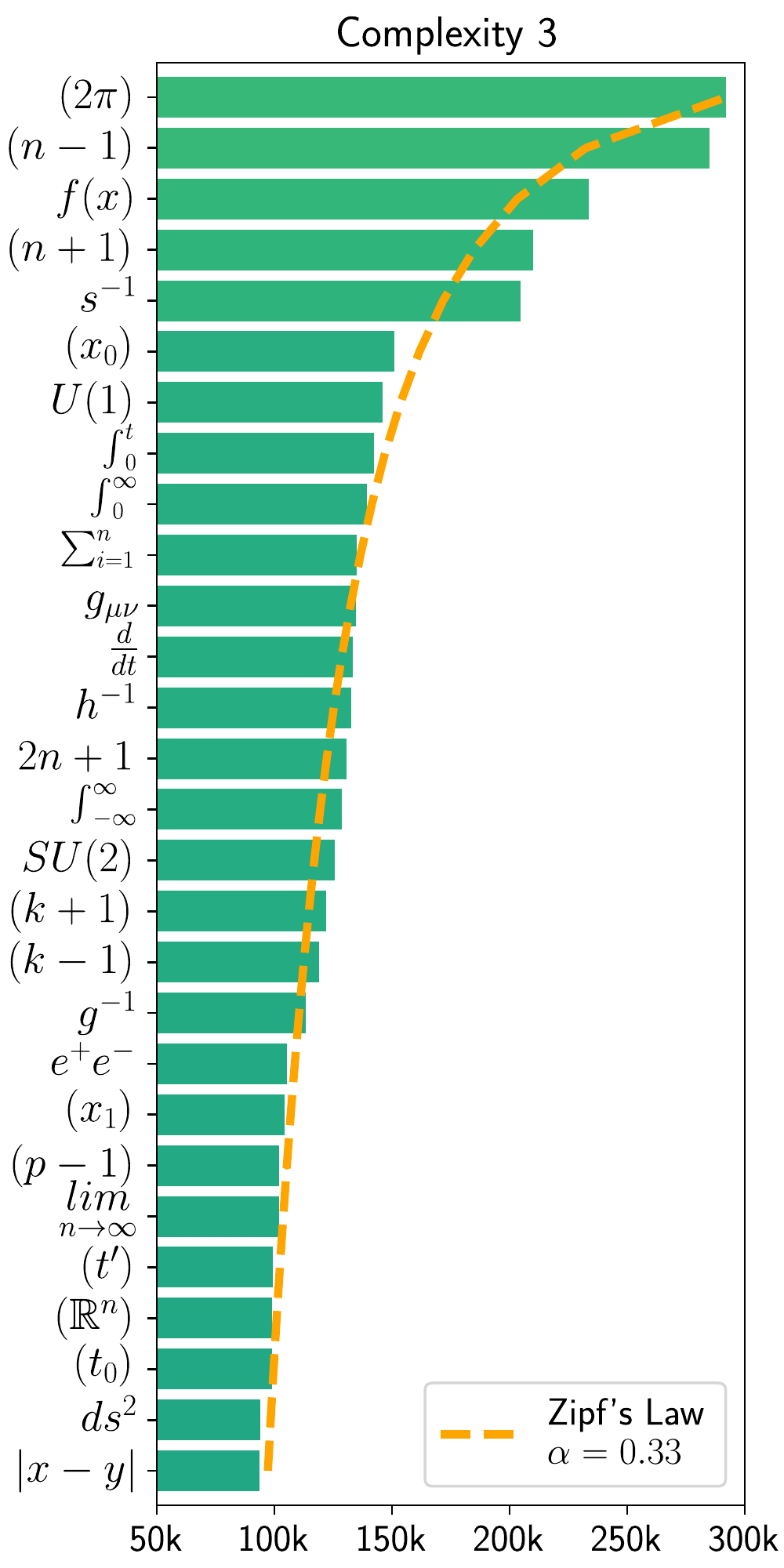}
	\end{subfigure}
  	\begin{subfigure}[b]{.196\textwidth}
		\centering
  		\includegraphics[width=\textwidth]{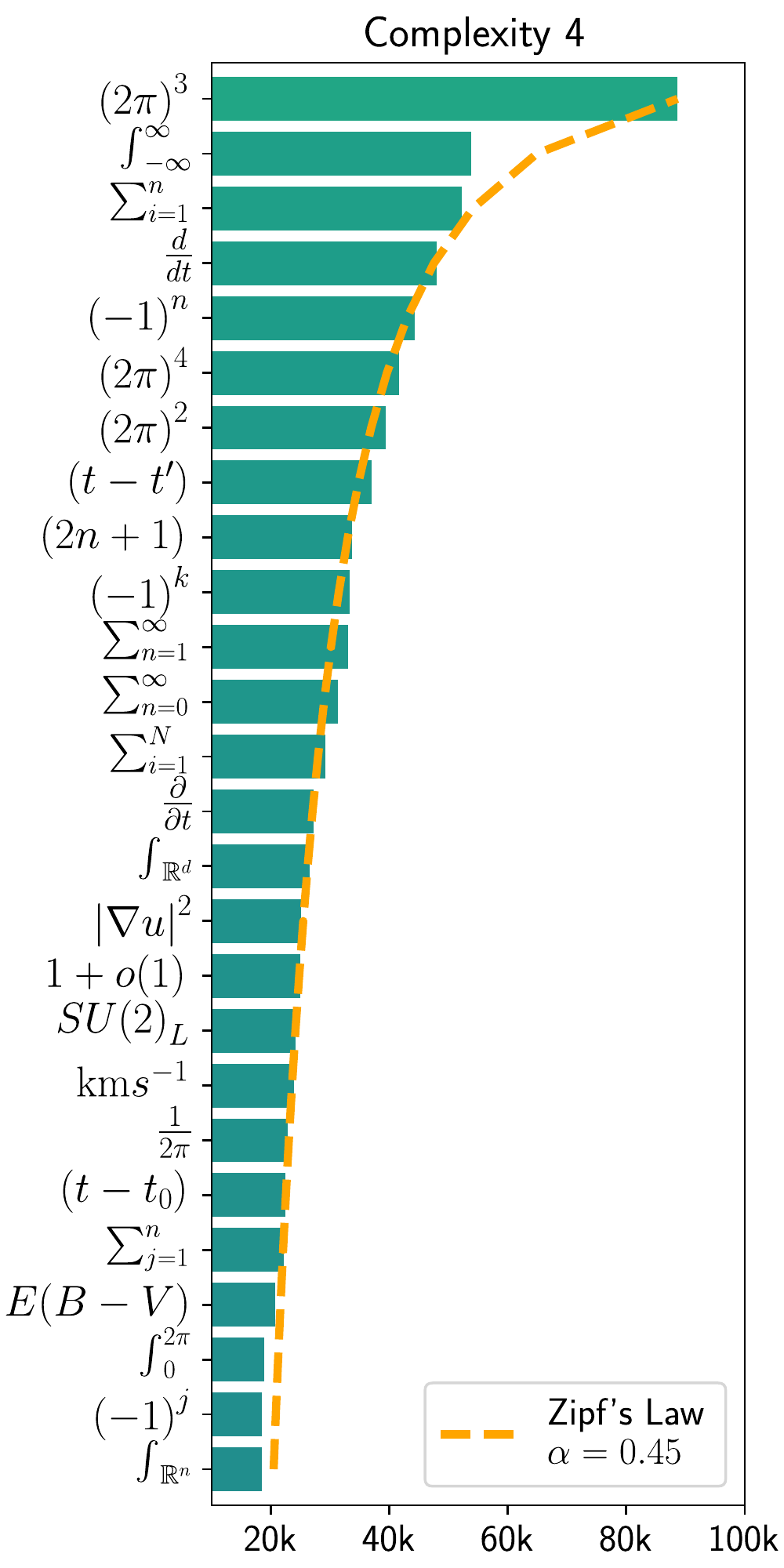}
	\end{subfigure}
	\begin{subfigure}[b]{.196\textwidth}
		\centering
		\includegraphics[width=\textwidth]{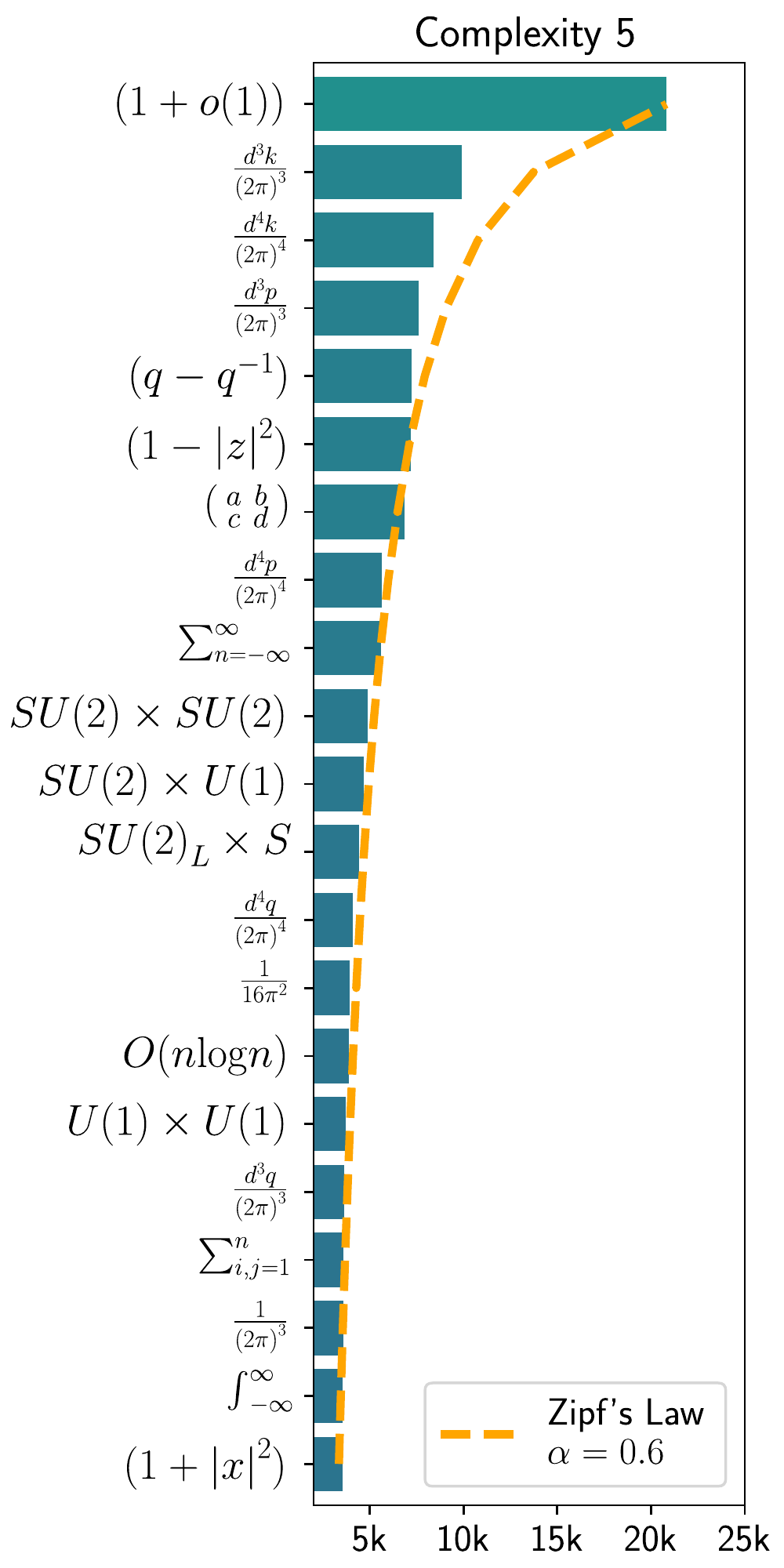}
	\end{subfigure}
	\vspace*{-0.3cm}
	\caption{Overview of the most frequent mathematical expressions in arXiv for complexities 1-5. The color gradient from yellow to blue represents the frequency in the dataset. Zipf's law~\eqref{eq:zipf-1} is represented by a dashed orange line.}
	\label{fig:arxiv-distributions}
	\vspace*{-0.3cm}
\end{figure*}
\subsection{Analyzing and Comparing Frequencies}
Figure~\ref{fig:arxiv-distributions} shows in detail the most frequently used mathematical expressions in arXiv for the complexities 1 to 5. 
The orange dashed line visible in all graphs represents the normal Zipf's law distribution from Equation~\eqref{eq:zipf-1}. 
We explore the total frequency values without any normalization. Thus, Equation~\eqref{eq:zipf-1} was multiplied by the highest frequency for each complexity level to fit the distribution. 
The plots in Figure~\ref{fig:arxiv-distributions} demonstrate that even though the parameter $\alpha$ varies 
between $0.35$ and $0.62$, the distributions in each complexity class also obey Zipf's law.

The plots for each complexity class contain some interesting fluctuations.
We can spot a set of five single identifiers that are most frequently used throughout arXiv:~$n$, $i$, $x$, $t$, and $k$. 
Even though the distributions follow Zipf's law accurately, we can explore that these five identifiers are proportionally more frequently used than other identifiers and clearly separate themselves above the rest (notice the large gap from $k$ to $a$).
All of the five identifiers are known to be used in a large variety of scenarios.
Surprisingly, one might expect that common pairs of identifiers would share comparable frequencies in the plots. However, typical pairs, such as $x$ and $y$, or $\alpha$ and $\beta$, possess a large discrepancy.

The plot of complexity two also reveals that two expressions are proportionally more often used 
than others:~$(x)$ and $(t)$. 
These two expressions appear more than three times as often in the corpus than any other expression of the same complexity.
On the other hand, the quantitative difference between $(x)$ and $(t)$ is negligible.
We may assume that arXiv's primary domain, physics, causes the quantitative disparity between $(x)$, $(t)$, and the other tokens.
The primary domain of the dataset becomes more clearly visible for higher complexities, such as $SU(2)$ (C3\footnote{We refer to a given complexity $n$ with C$n$, i.e., C3 refers to complexity 3.}) or $kms^{-1}$ (C4).

\begin{figure*}[htp]
	\centering
	\begin{subfigure}[b]{.20\textwidth}
		\centering
  		\includegraphics[height=9cm]{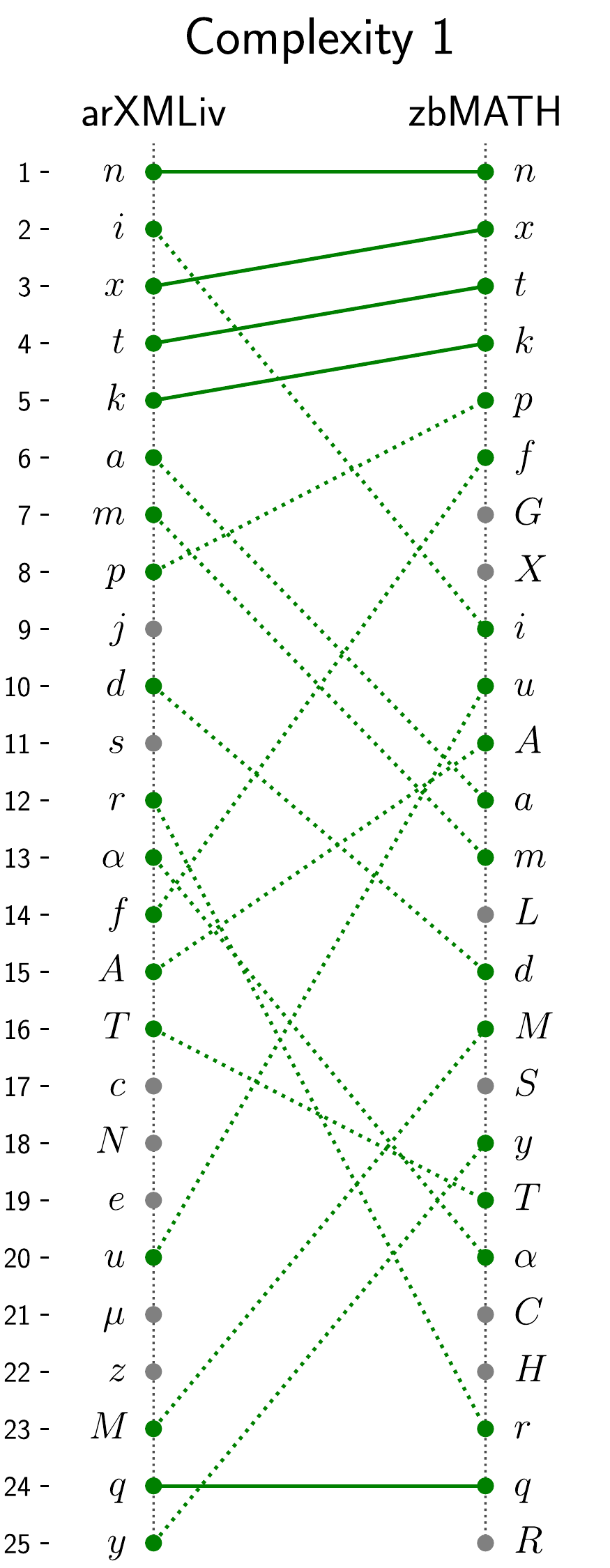}
	\end{subfigure}
	\hfill
  	\begin{subfigure}[b]{.22\textwidth}
		\centering
  		\includegraphics[height=9cm]{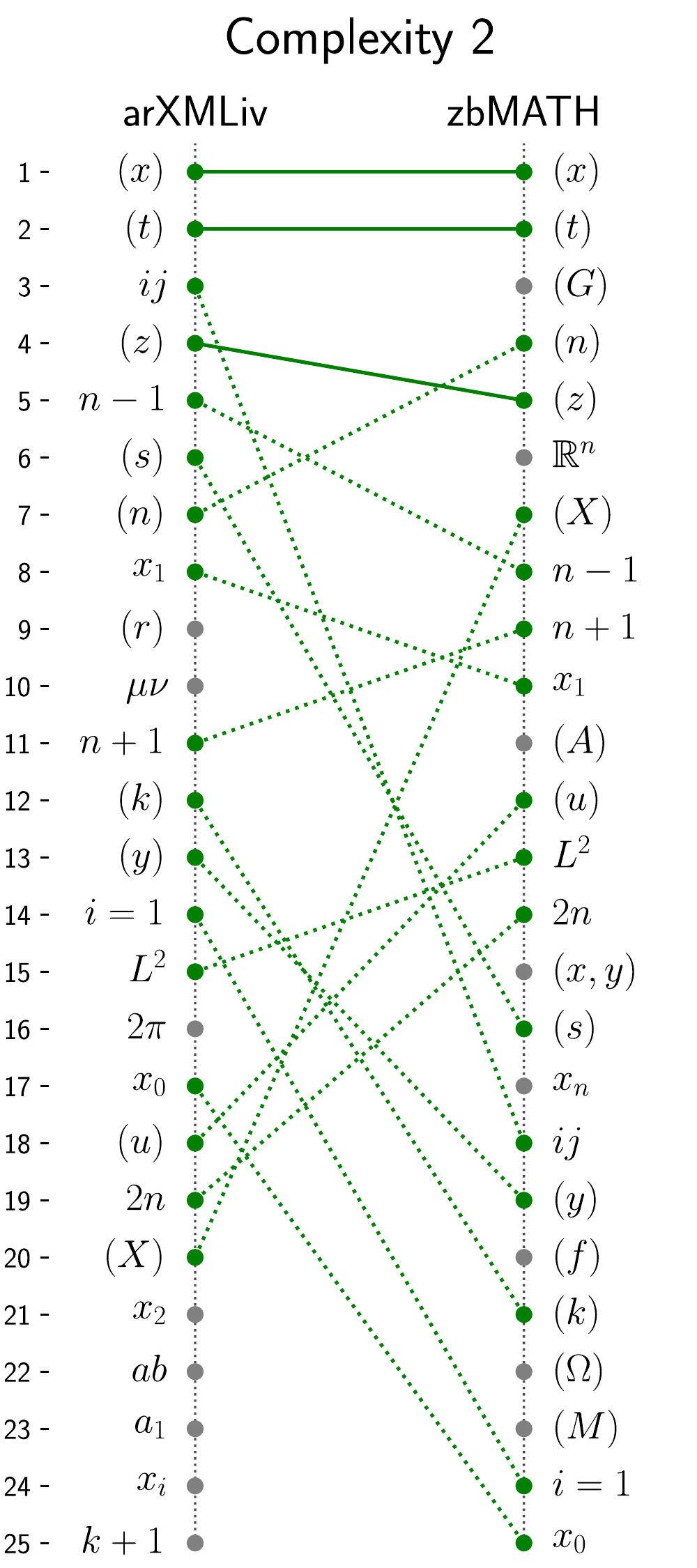}
	\end{subfigure}
	\hfill
  	\begin{subfigure}[b]{.26\textwidth}
		\centering
  		\includegraphics[height=9cm]{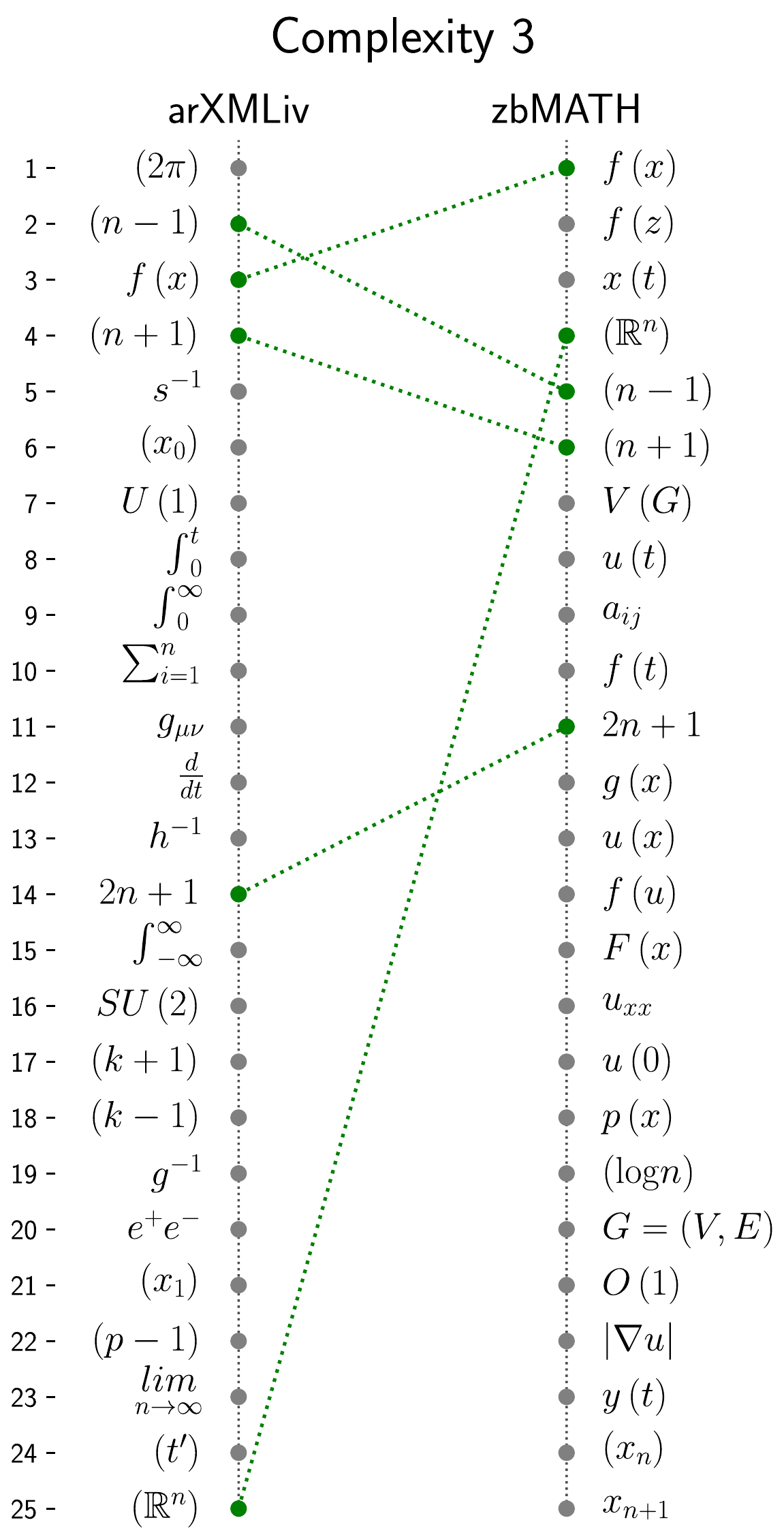}
	\end{subfigure}
	\hfill
  	\begin{subfigure}[b]{.3\textwidth}
		\centering
  		\includegraphics[height=9cm]{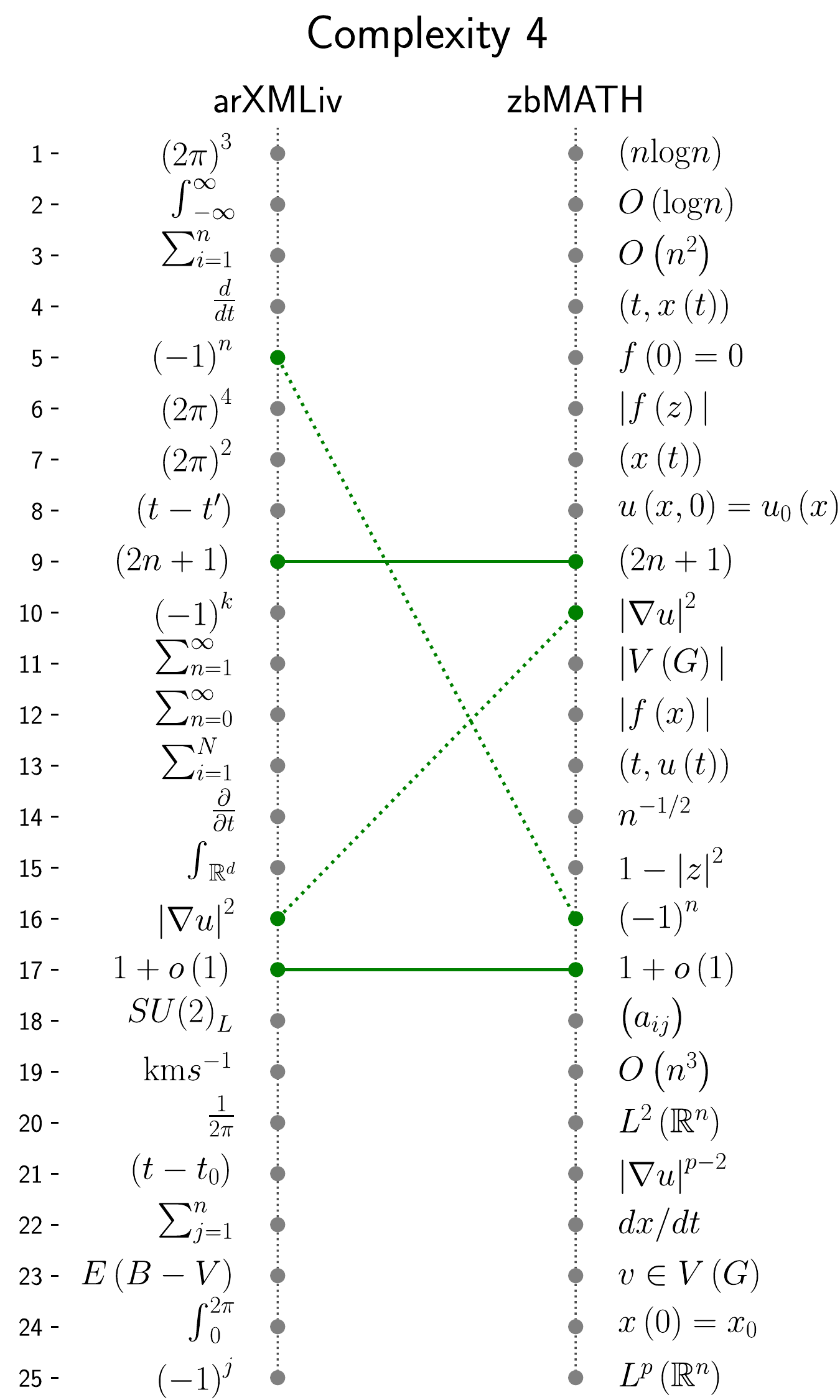}
	\end{subfigure}
	\vspace*{-0.35cm}
	\caption{The top-25 most frequent expressions in arXiv (left) and zbMATH (right) for complexities 1-4. A line between both sets indicates a matching set. Bold lines indicate that the matches share a similar rank (distance of 0 or 1).}
	\label{fig:dist-comparison}
	\vspace*{-0.4cm}
\end{figure*}

Another surprising property of arXiv is that symmetry groups, such as $SU(2)$, appear to play an essential role in the majority of articles on arXiv, see $SU(2)$ (C3), $SU(2)_L$ (C4), and $SU(2) \times SU(2)$ (C5), among others.
The plots of higher complexities\footnote{More plots showing higher complexities are available at \url{https://github.com/ag-gipp/FormulaCloudData}}, which we do not show here, made this even more noticeable. 
Given a complexity of six, for example, the most frequently used expression was $SU(2)_L \times SU(2)_R$, and for a complexity of seven it was $SU(3) \times SU(2) \times U(1)$.
Given a complexity of eight, ten out of the top-12 expressions were from symmetry group calculations.

It is also worthwhile to compare expressions among different levels of complexities.
For instance, $(x)$ and $(t)$ appeared almost six million times in the corpus, but $f(x)$ (at position three in C3) was the only expression which contained one of these most common expressions. 
Note that subexpressions of variations, such as $(x_0)$, $(t_0)$, or $(t-t')$, do not match the expression of complexity two.
This may imply that $(x)$, and especially $(t)$, appear in many different scenarios. 
Further, we can examine that even though $(x)$ is a part of $f(x)$ in only approximately 3\% of all cases, it is still the most likely combination.
These results are especially useful for recommendation systems that make use of math as input.
Moreover, plagiarism detection systems may also benefit from such a knowledge base.
For instance, it might be evident that $f(x)$ is a very common expression, but for automatic systems that work on a large scale, it is not clear whether duplicate occurrences of $f(x)$ or $\Xi(x)$ should be scored differently, e.g., in the case of plagiarism detection.

Figure~\ref{fig:arxiv-distributions} shows only the most frequently occurring expressions in arXiv.
Since we already explored a bias towards physics formulae in arXiv, it is worth comparing the expressions present within both datasets. Figure~\ref{fig:dist-comparison} compares the 25-top expressions for the complexities one to four. In zbMATH, we discovered that computer science and graph theory appeared as popular topics, see for example $G=(V,E)$ (in C3 at position 20) and the Bachmann-Landau notations in $O(\log n)$, $O(n^2)$, and $O(n^3)$ (C4 positions 2, 3, and 19).

\renewcommand{\arraystretch}{1.1}
\begin{table*}[tb]
	\centering
\begin{tabular}{*{4}{cc|}cc}

		\hline		
		\multicolumn{2}{c|}{C3} & \multicolumn{2}{c|}{C4} & \multicolumn{2}{c|}{C5} & \multicolumn{2}{c|}{C6} & \multicolumn{2}{c}{C7} \\\hline
		114.84 & $(n!)$ &
		129.44 & $i,j = 1, \ldots, n$ &
		119.21 & $\operatorname{Gal}\!\left(\overline{\mathbb{Q}} /  \mathbb{Q}\right)$ &
		110.83 & $( 1 + \left| z \right|^2 )^{\alpha}$ &
		98.72 & $\operatorname{div}\!\left( \left| \nabla u\right|^{p-2} \nabla u\right)$
		\\
		
		108.85 & $\phi^{-1}$ &
		108.52 & $x_{ij}$ &
		112.55 & $\left| f(z) \right|^p$ &
		105.69 & $f\!\left( re^{i\theta} \right)$ &
		\multicolumn{2}{c}{--}
		\\
		
		100.19 & $z^{n-1}$ &
		108.50 & $\dot{x} = A(t)x$ &
		110.52 & $\left( 1 + \left| x \right|^2 \right)$ &
		94.14 & $f(z) = z + \sum_{n=2}^{\infty} a_n z^n$&
		\multicolumn{2}{c}{--}
		\\
		
		100.06 & $(c_n)$ &
		106.66 & $|x-x_0|$ &
		109.19 & $\left| f(x) \right|^p$ &
		92.33 & $\left( \left| \nabla u\right|^{p-2} \nabla u\right)$ &
		\multicolumn{2}{c}{--}
		\\
		
		100.05 & $B(G)$ &
		105.52 & $S^{2n+1}$ &
		106.22 & $|\nabla u |^2 dx$ &
		87.27 & $\left(\log n / \log \log n \right)$ &
		\multicolumn{2}{c}{--}
		\\
		
		99.87 & $\log_2 n$ &
		104.91 & $L^2\!\left( \mathbb{R}^2 \right)$ &
		102.86 & $n(n-1)/2$ &
		78.54 & $O\,(n \log^2 n )$&
		\multicolumn{2}{c}{--}
		\\
		
		99.65 & $\xi\,(x)$ &
		103.70 & $\dot{x} = Ax + Bu$ &
		101.40 & $O(n^{-1})$ &
		\multicolumn{2}{c|}{--} &
		\multicolumn{2}{c}{--}
		\\\hline
	\end{tabular}%
	\vspace*{0cm}
	\caption{Top $s(t,D)$ scores, where $D$ is the set of all zbMATH documents with a minimum document frequency of 200, maximum document frequency of 500k, and a minimum complexity of 3.}
	\label{tab:top-tfidf-zbm}
	\vspace*{-0.9cm}
\end{table*}

From Figure~\ref{fig:dist-comparison}, we can also deduce useful information for MathIR tasks which focus on semantic information. Current semantic extraction tools~\cite{Schubotz2017} or \LaTeX{} parsers~\cite{Schubotz2018c} still have difficulties distinguishing \emph{multiplications} from \emph{function calls}. For example as mentioned before, \LaTeXML~\cite{LaTeXML} adds an \textit{invisible times} character between $f(x)$ rather than a \textit{function application}. Investigating the most frequently used terms in zbMATH in Table~\ref{fig:dist-comparison} reveals that $u$ is most likely considered to be a function in the dataset:~$u(t)$ (rank 8), $u(x)$ (rank 13), $u_{xx}$ (rank 16), $u(0)$ (rank 17), $\left|\nabla u \right|$ (rank 22). Manual investigations of extended lists reveal even more hits:~$u_0(x)$ (rank 30), $-\Delta u$ (rank 32), and $u(x,t)$ (rank 33).
Since all eight terms are among the most frequent 35 entries in zbMATH, it implies that $u$ can most likely be considered to imply a function in zbMATH. 
Of course, this does not imply that $u$ must always be a function in zbMATH (see $f(u)$ on rank 14 in C3), but this allows us to exploit probabilities for improving MathIR performance. 
For instance, if not stated otherwise, $u$ could be interpreted as a function by default, which could help increase the precision of the aforementioned tools.

Figure~\ref{fig:dist-comparison} also demonstrates that our two datasets diverge for increasing complexities.
Hence, we can assume that frequencies of less complex formulae are more topic-independent.
Conversely, the more complex a math formula is, the more context-specific it is.
In the following, we will further investigate this assumption by applying TF-IDF rankings on the distributions.

\vspace{-.2cm}
\section{Relevance Ranking for Formulae}
Zipf's law %
encourages the idea of scoring the relevance of words according to their number of occurrences in the corpus and in the documents. 
The family of BM25 ranking functions based on TF-IDF scores are still widely used in several retrieval systems~\cite{RobertsonZ09,BeelGLB16}.
Since we demonstrated that mathematical formulae (and their subexpressions) obey Zipf's law in large scientific corpora, it appears intuitive to also use TF-IDF rankings, such as a variant of BM25, to calculate their relevance.
In its original form~\cite{RobertsonZ09}, \textit{Okapi BM25} was calculated as follows
\begin{equation}\label{eq:bm25}
\operatorname{bm25}(t,d) := \frac{\left( k + 1 \right) \operatorname{IDF}(t) \operatorname{TF}(t,d)}{\operatorname{TF}(t,d) + k \left( 1-b + \frac{b |d|}
{\AVGsub{DL}}
\right)},
\end{equation}
where $\operatorname{TF}\,(t,d)$ is the term frequency of $t$ in the document $d$, $|d|$ the length of the document $d$ (in our case, the number of subexpressions), 
$\AVGsub{DL}$
the average length of the documents in the corpus (see Table~\ref{tab:summary}), and $\operatorname{IDF}\,(t)$ is the inverse document frequency of $t$, defined as
\begin{equation}\label{eq:bm25-idf}
\vspace*{-0.2cm}
\operatorname{IDF}(t) := \log \frac{N - n(t) + \tfrac{1}{2}}{n(t) + \tfrac{1}{2}},
\end{equation}
where $N$ is the number of documents in the corpus and $n(t)$ the number of documents which contain the term $t$. By adding $\tfrac{1}{2}$, we avoid $\log 0$ and division by $0$.
The parameters $k$ and $b$ are free, with $b$ controlling the influence of the normalized document 
length and $k$ controlling the influence of the term frequency on the final score.
For our experiments, we chose the standard value $k=1.2$ and a high impact factor of the normalized document length via $b=0.95$.

\begin{figure*}[tp]
	\centering
	\vspace*{0.15cm}
	\begin{subfigure}[b]{.19\textwidth}
		\centering
  		\includegraphics[height=6.6cm]{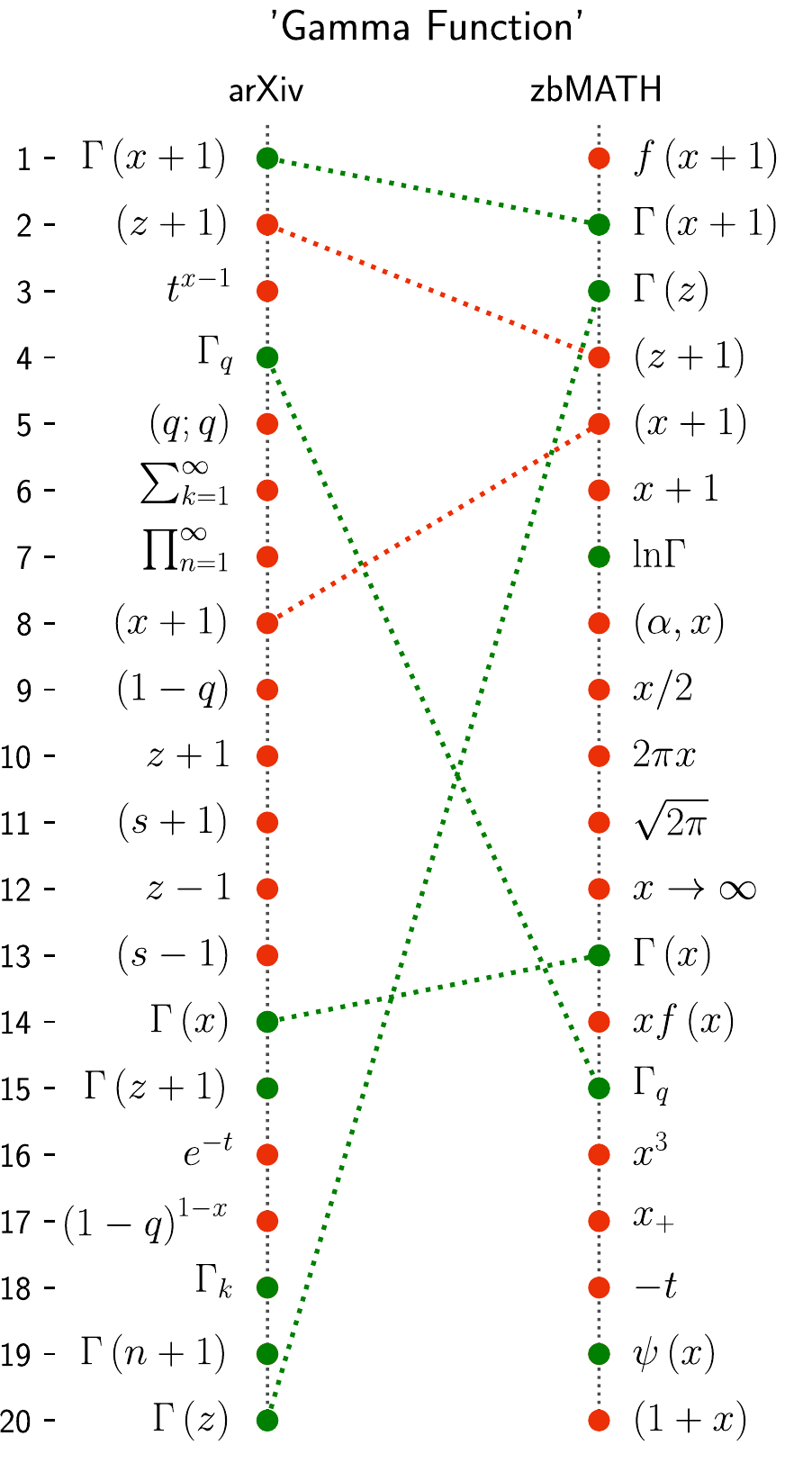}
	\end{subfigure}
	\hfill
  	\begin{subfigure}[b]{.24\textwidth}
		\centering
  		\includegraphics[height=6.6cm]{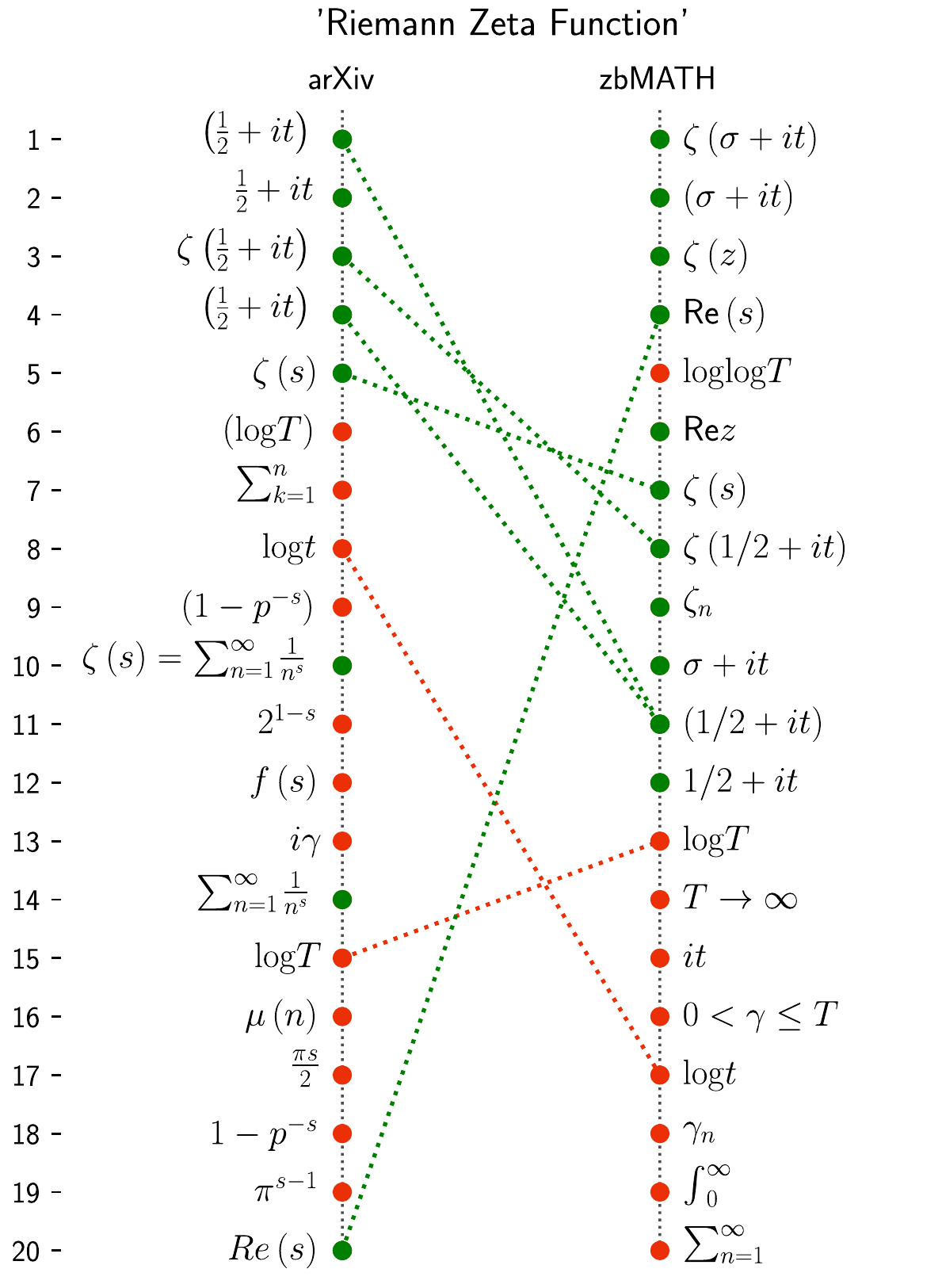}
	\end{subfigure}
	\hfill
  	\begin{subfigure}[b]{.18\textwidth}
		\centering
  		\includegraphics[height=6.6cm]{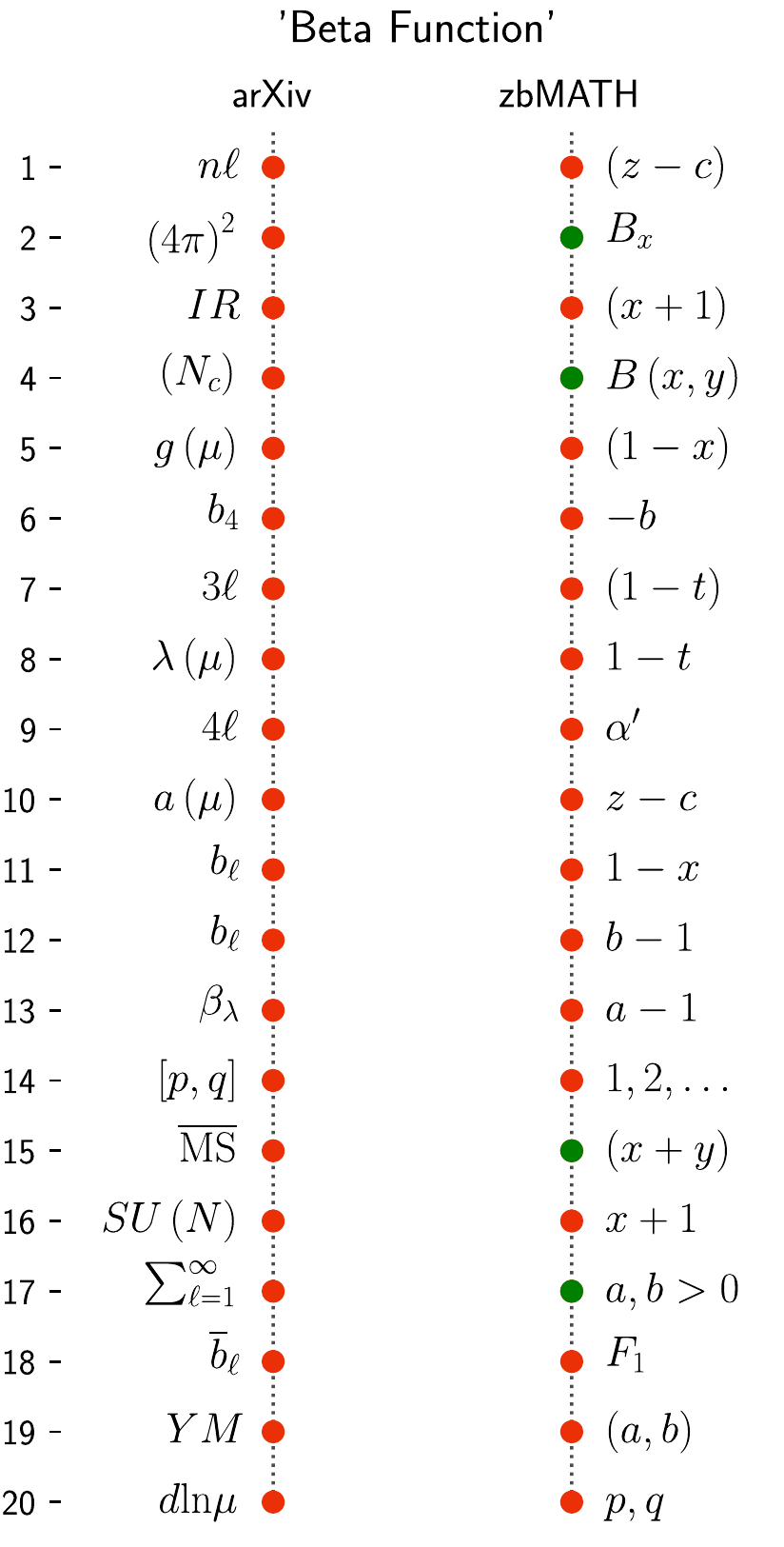}
	\end{subfigure}
	\hfill
  	\begin{subfigure}[b]{.18\textwidth}
		\centering
  		\includegraphics[height=6.6cm]{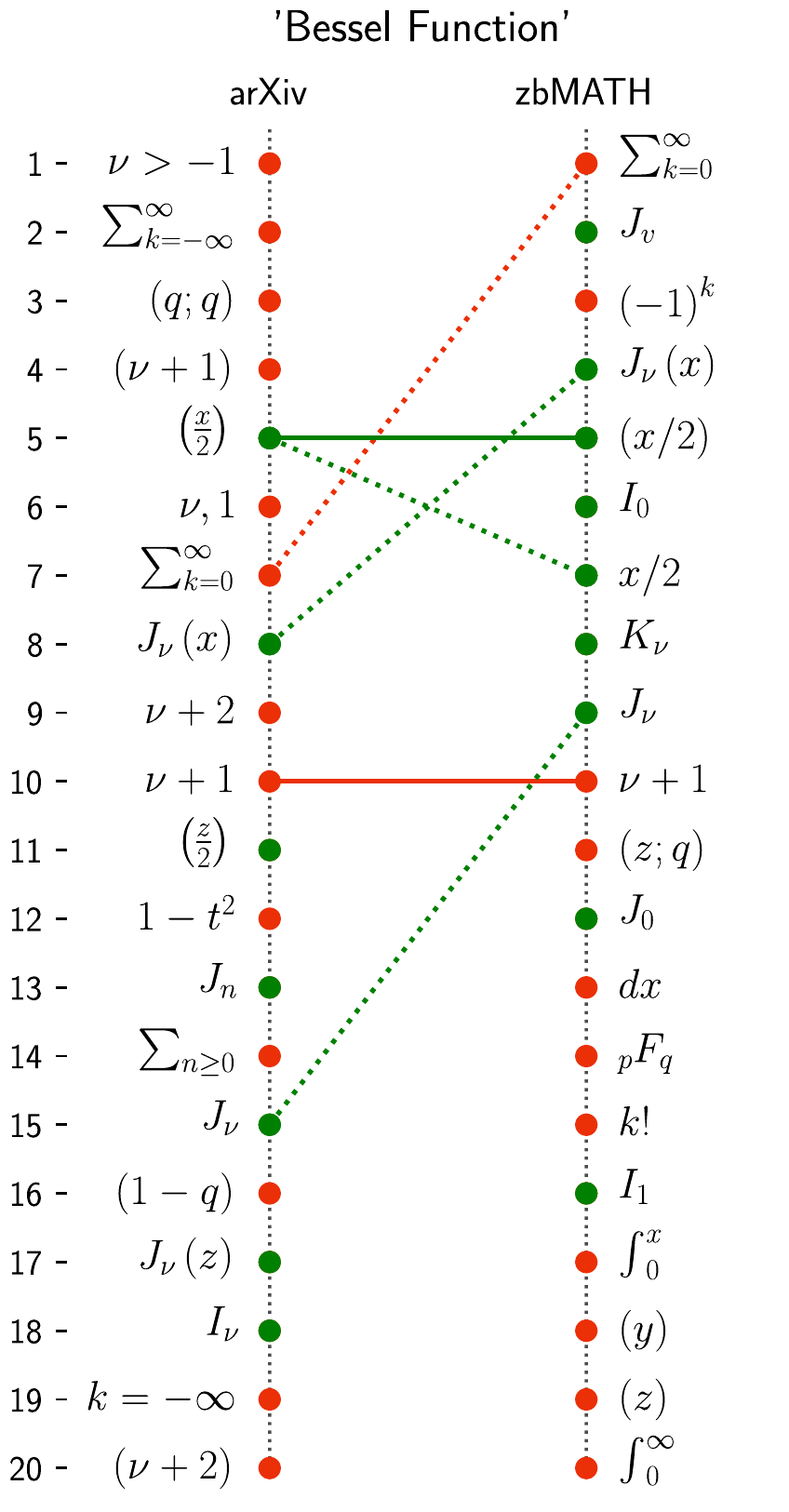}
	\end{subfigure}
	\hfill
  	\begin{subfigure}[b]{.18\textwidth}
		\centering
  		\includegraphics[height=6.6cm]{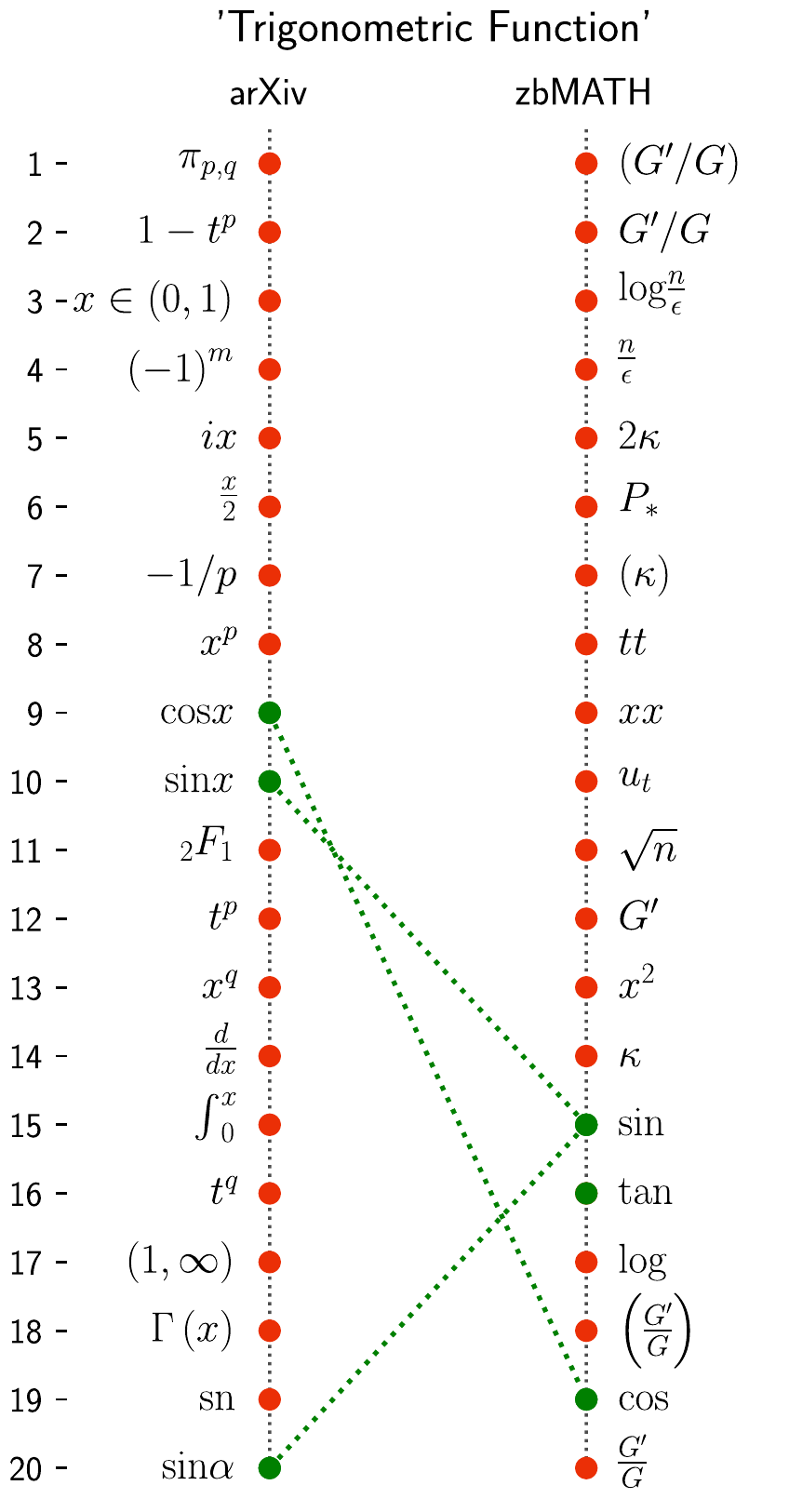}
	\end{subfigure}
	\vspace*{-0.45cm}
	\caption{Top-20 ranked expressions retrieved from a topic-specific subset of documents $D_q$. The search query $q$ is given above the plots. Retrieved formulae are annotated by a domain expert with green dots for relevant and red dots for non-relevant hits. A line is drawn if a hit appears in both result sets. The line is colored in green when the hit was marked as relevant.}
	\label{fig:tfidf-all}
	\vspace*{-0.4cm}
\end{figure*}

As a result of our subexpression extraction algorithm, we generated a bias towards low complexities.
Moreover, longer documents generally consist of more complex expressions.
As demonstrated in Section~\ref{sec:data-wrangling}, a document that only consists of the single expression $\StandardJacobi$, i.e., the document had a length of one, would generate eight subexpressions, i.e., it results in a document length of eight.
Thus, we modify the BM25 score in Equation~\eqref{eq:bm25} to emphasize higher complexities and longer documents. 
First, the average document length is divided by the average complexity 
$\AVGsub{C}$
in the corpus that is used (see Table~\ref{tab:summary}), and we calculate the reciprocal of the document length normalization to emphasize longer documents.

Moreover, in the scope of a single document, we want to emphasize expressions that do not appear frequently in this document, but are the most frequent among their level of complexity. 
Thus, less complex expressions are ranked more highly if the document overall is not very complex. 
To achieve this weighting, we normalize the term frequency of an expression $t$ according to its complexity $c(t)$ and introduce an inverse term frequency according to all expressions in the document
\vspace*{-0.05cm}
\begin{equation}\label{eq:itf}
\operatorname{ITF}(t,d) := \log \frac{|d| - \operatorname{TF}(t,d) + \tfrac{1}{2}}{\operatorname{TF}(t,d) + \tfrac{1}{2}}.
\vspace*{-0.05cm}
\end{equation}
Finally, we define the score $\operatorname{s}(t,d)$ of a term $t$ in a document $d$ as
\vspace*{0.1cm}
\begin{equation}\label{eq:bm25-new}
s(t,d) := \frac{\left( k + 1 \right) \operatorname{IDF}(t) \operatorname{ITF}(t,d) \operatorname{TF}(t,d)}{\underset{t' \in d|_{c(t)}}{\operatorname{max}}\! \operatorname{TF}(t',d) + k \left( 1-b + \frac{b 
\AVGsub{DL}
}
{|d| 
\AVGsub{C}
} 
\right)
}.
\vspace*{-0.1cm}
\end{equation}
The TF-IDF ranking functions and the introduced $\operatorname{s}\,(t,d)$ are used to retrieve relevant documents for a given search query. However, we want to retrieve relevant subexpressions over a set of documents. Thus, we define the score of a formula (mBM25) over a set of documents as the maximum score over all documents
\vspace*{-0.05cm}
\begin{equation}\label{eq:ranking}
\operatorname{mBM25}(t,D) := \underset{d \in D}{\operatorname{max}} \operatorname{s}\,(t,d),
\end{equation}
where $D$ is a set of documents.
We used \textit{Apache Flink}~\cite{HueskeW19} to count the expressions and process the calculations. Thus, our implemented system scales well for large corpora.

Table~\ref{tab:top-tfidf-zbm} shows the top-7 scored expressions, where $D$ is the entire zbMATH dataset. 
The retrieved expressions can be considered as meaningful and real-world examples of MOIs, since most expressions are known for specific mathematical concepts, such as $\operatorname{Gal}(\overline{\mathbb{Q}}/\mathbb{Q})$, which refers to the Galois group of $\overline{\mathbb{Q}}$ over $\mathbb{Q}$, or $L^2(\mathbb{R}^2)$, which refers to the $L^2$-space (also known as \textit{Lebesgue space}) over $\mathbb{R}^2$.
However, a more topic-specific retrieval algorithm is desirable. 
To achieve this goal, we (i) retrieved a topic-specific subset of documents $D_{q} \subset D$ for a given textual search query $q$, and (ii) calculated the scores of all expressions in the retrieved documents. To generate $D_q$, we indexed the text sources of the documents from arXiv and zbMATH via elasticsearch (ES)\footnote{\url{https://github.com/elastic/elasticsearch} [Accessed Sep.~2019]. We used version 7.0.0} and performed the pre-processing steps:~filtering stop words, stemming, and ASCII-folding\footnote{This means that non-ASCII characters are replaced by their ASCII counterparts or will be ignored if no such counterpart exists.}. 
Table~\ref{tab:tfidf-settings} summarizes the settings we used to retrieve MOIs from a topic-specific subset of documents $D_q$. 
We also set a minimum hit frequency according to the number of retrieved documents an expression appears in. 
This requirement filters out uncommon notations. 
\renewcommand{\arraystretch}{1}
\begin{wrapfigure}{R}{0.27\textwidth}
\vspace*{-0.5cm}
\hspace*{-0.15cm}
\centering
\begin{tabular}{r|cc}
	& arXiv & zbMATH \\\hline
	Retrieved Doc. & 40 & 200 \\
	Min.~Hit Freq. & 7 & 7 \\
	Min.~DF & 50 & 10 \\
	Max.~DF & 10k & 10k \\
\end{tabular}
\vspace*{-0.38cm}
\makeatletter\def\@captype{table}\makeatother %
\caption{Settings for the retrieval experiments.}
\label{tab:tfidf-settings}
\vspace*{-0.45cm}
\end{wrapfigure}

Figure~\ref{fig:tfidf-all} shows the results for five search queries.
We asked a domain expert from the National Institute of Standards and Technology (NIST) to annotate the results as related (shown as green dots in Figure~\ref{fig:tfidf-all}) or non-related (red dots).
We found that the results range from good performances (e.g., for the Riemann zeta function) to bad performances (e.g., beta function).
For instance, the results for the Riemann zeta function are surprisingly accurate, since we could discover that parts of Riemann's hypothesis\footnote{Riemann proposed that the real part of every non-trivial zero of the Riemann zeta function is $1/2$. If this hypothesis is correct, all the non-trivial zeros lie on the critical line consisting of the complex numbers $1/2 + it$.} were ranked highly throughout the results (e.g., $\zeta(\frac{1}{2}+it)$).
On the other hand, for the beta function, we retrieved only a few related hits, of which only one had a strong connection to the beta function $B(x,y)$.
We observed that the results were quite sensitive to the chosen settings (see Table~\ref{tab:tfidf-settings}).
For instance, according to the beta function, the minimum hit frequency has a strong effect on the results, since many expressions are shared among multiple documents.
For arXiv, the expressions $B(\alpha, \beta)$ and $B(x,y)$ only appear in one document of the retrieved 40. However, decreasing the minimum hit frequency would increase noise in the results.

\renewcommand{\arraystretch}{1.2}
\begin{table*}
\vspace*{-0.5cm}
	\centering
\begin{tabular}{*{6}{rc|}|c|c}
	\multicolumn{14}{c}{Riemann Zeta Function} \\[-0.06cm] \hline
	\multicolumn{2}{c|}{C1} & \multicolumn{2}{c|}{C2} & \multicolumn{2}{c|}{C3} & \multicolumn{2}{c|}{C4} & \multicolumn{2}{c|}{C5} & \multicolumn{2}{c||}{C6} & TF-IDF & mBM25 \\\hline
	15,051 & $n$ & 4,663 & $(s)$ & 1,456 & $\zeta(s)$         & 349 & $(\frac{1}{2} + it)$  & 203 & $\zeta(\frac{1}{2} + it)$ & 105 & $\left|\zeta(1/2 + it)\right|$ & $\zeta (s)$ & $\zeta\,(1/2 + it)$ \\
11,709 & $s$ & 2,460 & $(x)$ &   340 & $\sigma + it$      & 232 & $(1/2 + it)$          & 166 & $\zeta(1/2 + it)$         & 88 & $\left|\zeta(\frac{1}{2} + it)\right|$  & $\zeta ( 1/2 + it )$ & $(1/2 + it)$ \\
 9,768 & $x$ & 2,163 & $(n)$ &   310 & $\sum_{n=1}^\infty$& 195 & $(\sigma + it)$       & 124 & $\zeta(\sigma + it)$      & 81 & $\left|\zeta(\sigma + it)\right|$ & $(1/2 + it)$ & $(\frac{1}{2} + it)$\\
 8,913 & $k$ & 1,485 & $(t)$ &   275 & $(\log T)$ & 136   & $\frac{1}{2} + it$        & 54 & $\zeta(1 + it)$            & 32 & $\left|\zeta(1 + it)\right|$ & $\frac{1}{2} + it$ & $\zeta\,(\frac{1}{2} + it)$ \\
 8,634 & $T$ & 1,415 & $it$  &   264 & $1/2 + it$ & 97    & $s = \sigma + it$           & 44 & $\zeta(2n + 1)$            & 22 & $\left|\zeta(+it)\right|$ & $(\frac{1}{2} + it)$ & $(\sigma + it)$ \\\hline
	\end{tabular}
\vspace*{0.1cm}
	\setlength{\tabcolsep}{4pt}%
\begin{tabular}{*{6}{rc|}|c|c}
	\multicolumn{14}{c}{Eigenvalue} \\[-0.06cm] \hline
	\multicolumn{2}{c|}{C1} & \multicolumn{2}{c|}{C2} & \multicolumn{2}{c|}{C3} & \multicolumn{2}{c|}{C4} & \multicolumn{2}{c|}{C5} & \multicolumn{2}{c||}{C6} & TF-IDF & mBM25 \\\hline
	45,488 & $n$       & 12,515 & $(x)$         & 686 & $-\Delta u$            & 218 & $\left|\nabla u\right|^{p-2}$  & 139 & $\left|\nabla u\right|^{p-2} \nabla u$ & 137 & $\left(\left|\nabla u\right|^{p-2} \nabla u\right)$ \TT & $A x = \lambda B x$ & $- \operatorname{div}\left(\left|\nabla u\right|^{p-2} \nabla u\right)$ \\
43,090 & $x$       &  6,598 & $(t)$         & 555 & $(n-1)$                & 218 & $-\Delta_p u$                  &  68 & $-d^2 / dx^2$                          & 35  & $-(py')^{'}$ & $- \Delta p$ & $\operatorname{div}\left(\left|\nabla u\right|^{p-2} \nabla u\right)$ \\
37,434 & $\lambda$ &  4,377 & $\lambda_1$   & 521 & $\left|\nabla u\right|$& 133 & $W_0^{1,p}(\Omega)$            &  51 & $A = (a_{ij})$                         & 26  & $(\left|u'\right|^{p-2}u')$ & $P( \lambda )$ & $p = \frac{N+2}{N-2}$\\
35,302 & $u$       &  2,787 & $(\Omega)$    & 512 & $a_{ij}$               & 127 & $\left|\nabla u\right|^2$      &  46 & $-\frac{d^2}{dx^2}$                    & 18  & $(\phi_p (u'))^{'}$  & $\lambda_{k+1}$ & $\left( \phi_p \left( u' \right) \right) '$\\
\rule{0pt}{11pt}22,460 & $t$       &  2,725 & $\mathbb{R}^n$& 495 & $u(x)$                 &  97 & $(a_{ij})$                     &  45 & $u \in W_0^{1,p}(\Omega)$              & 18  & $\int_{\Omega} \left| \nabla u \right|^2 dx$ \TB & $\lambda_1 > 0$ & $\lambda \in (0, \lambda^*)$
	\\\hline
	\end{tabular}
	\caption{The top-5 frequent mathematical expressions in the result set of zbMATH for the search queries `Riemann Zeta Function' (top) and `Eigenvalue' (bottom) grouped by their complexities (left) and the hits reordered according to their relevance scores (right). The TF-IDF score was calculated with normalized term frequencies.}
	\label{tab:zbmath-search-engine}
\vspace*{-0.8cm}
\end{table*}

Even though we asked a domain expert to annotate the results as relevant or not, there is still plenty of room for discussion.
For instance, $(x+y)$ (rank 15 in zbMATH, `Beta Function') is the argument of the gamma function $\Gamma(x+y)$ that appears in the definition of the beta function \cite[(5.12.1)]{NIST:DLMF} $B(x,y):=\Gamma(x)\Gamma(y)/\Gamma(x+y)$.
However, this relation is weak at best, and thus might be considered as not related.
Other examples are $\mathrm{Re} z$ and $\mathrm{Re} (s)$, which play a crucial role in the scenario of the Riemann hypothesis (all non-trivial zeroes have $\mathrm{Re} (s) = \frac{1}{2}$). Again, this connection is not obvious, and these expressions are often used in multiple scenarios. 
Thus, the domain expert did not mark the expressions as being related.

Considering the differences in the documents, it is promising to have observed a relatively high number of shared hits in the results.
Further, we were able to retrieve some surprisingly good insights from the results, such as extracting the full definition of the Riemann zeta function \cite[(25.2.1)]{NIST:DLMF} $\zeta(s) := \sum_{n=1}^{\infty} \frac{1}{n^s}$.
Even though a high number of shared hits seem to substantiate the reliability of the system, there were several aspects that affected the outcome negatively, from the exact definition of the search queries to retrieve documents via ES, to the number of retrieved documents, the minimum hit frequency, and the parameters in mBM25.

\section{Applications}
The presented results are beneficial for a variety of use-cases. In the following, we will demonstrate and discuss several of the applications that we propose.

\begin{table}[tb]
	\centering
	\begin{tabular}{lrr|lrr}
	\multicolumn{3}{c}{Auto-completion for `$E = m$'} & \multicolumn{3}{c}{Suggestions for `$E = \{m,c\}$'} \\\hline
	Sug.~Expression & TF & DF & Sug.~Expression & TF & DF \\\hline
	$E = mc^2$ & 558 & 376 					& $E = mc^2$  & 558 & 376  \\
	$E = m\cosh \theta$ & 23 & 23 			& $E = \gamma mc^2$ & 39 & 38  \\
	$E = mv_0$ & 7 & 7 						& $E = \gamma m_e c^2$ & 41 & 36 \\
	$E = m/\sqrt{1-\dot{q}^2}$ & 12 & 6		& $E = m \cosh \theta$ & 23 & 23  \\
	$E = m/\sqrt{1-\beta^2}$ & 10 & 6		& $E = -mc^2$ & 35 & 17  \\
	$E = mc^2\gamma$ & 6 & 6				& $E = \sqrt{m^2c^4 + p^2c^2}$ & 10 & 8  \\
	\hline
	\end{tabular}
	\caption{Suggestions to complete `$E = m$' and `$E = \{m,c\}$' (the right-hand side contains $m$ and $c$) with term and document frequency based on the distributions of formulae in arXiv.}
	\label{tab:recommendation}
	\vspace*{-1cm}
\end{table}

\noindent
\textbf{Extension of zbMATH's Search Engine:} Formula search engines are often counterintuitive when compared to textual search, since the user must know how the system operates to enter a search query properly (e.g., does the system supports \LaTeX{} inputs?). 
Additionally, mathematical concepts can be difficult to capture using only mathematical expressions. 
Consider, for example, someone who wants to search for mathematical expressions that are related to eigenvalues.
A textual search query would only retrieve entire documents that require further investigation to find related expressions.
A mathematical search engine, on the other hand, is impractical since it is not clear what would be a fitting search query (e.g., $Av = \lambda v$?).
Moreover, formula and textual search systems for scientific corpora are separated from each other.
Thus, a textual search engine capable of retrieving mathematical formulae can be beneficial.
Also, many search engines allow for narrowing down relevant hits by suggesting filters based on the retrieved results. 
This technique is known as faceted search.
The zbMATH search engine also provides faceted search, e.g., by authors, or year.
Adding facets for mathematical expressions allows users to narrow down the results more precisely to arrive at specific documents.

Our proposed system for extracting relevant expressions from scientific corpora via mBM25 scores can be used to search for formulae even with textual search queries, and to add more filters for faceted search implementations. 
Table~\ref{tab:zbmath-search-engine} shows two examples of such an extension for zbMATH's search engine.
Searching for `Riemann Zeta Function' and `Eigenvalue' retrieved 4,739 and 25,248 documents from zbMATH, respectively.
Table~\ref{tab:zbmath-search-engine} shows the most frequently used mathematical expressions in the set of retrieved documents.
It also shows the reordered formulae according to a default TF-IDF score (with normalized term frequencies) and our proposed mBM25 score.
The results can be used to add filters for faceted search, e.g., show only the documents which contain $u \in W_0^{1,p}(\Omega)$.
Additionally, the search system now provides more intuitive textual inputs even for retrieving mathematical formulae.
The retrieved formulae are also interesting by themselves, since they provide insightful information on the retrieved publications. 
As already explored with our custom document search system in Figure~\ref{fig:tfidf-all}, the Riemann hypothesis is also prominent in these retrieved documents.

The differences between TF-IDF and mBM25 ranking illustrates the problem of an extensive evaluation of our system.
From a broader perspective, the hit $Ax = \lambda Bx$ is highly correlated with the input query `Eigenvalue'. On the other hand, the raw frequencies revealed a prominent role of $\operatorname{div}(\left|\nabla u\right|^{p-2} \nabla u)$. Therefore, the top results of the mBM25 ranking can also be considered as relevant. 

\noindent
\textbf{Math Notation Analysis:} A faceted search system allows us to analyze mathematical notations in more detail. For instance, we can retrieve documents from a specific time period. 
This allows one to study the evolution of mathematical notation over time~\cite{HistoryNotation}, or for identifying trends in specific fields. 
Also, we can analyze standard notations for specific authors since it is often assumed that authors prefer a specific notation style which may vary from the standard notation in a field. 
\balance

\noindent
\textbf{Math Recommendation Systems:} The frequency distributions of formulae can be used to realize effective math recommendation tasks, such as type hinting or error-corrections. 
These approaches require long training on large datasets, but may still generate meaningless results, such as $G_i = \{ (x,y)\in \mathbb{R}^n :~x_i = x_i \}$~\cite{Yasunaga2019}. 
We propose a simpler system which takes advantage of our frequency distributions. 
We retrieve entries from our result database, which contain all unique expressions and their frequencies. 
We implemented a simple prototype that retrieves the entries via pattern matching.
Table~\ref{tab:recommendation} shows two examples. The left side of the table shows suggested autocompleted expressions for the query `$E\!=\!m$'. The right side shows suggestions for `$E\!=$', where the right-hand side of the equation should contain $m$ and $c$ in any order. 
A combination using more advanced retrieval techniques, such as similarity measures based on symbol layout trees~\cite{Zanibbi2016,DavilaZ17}, would enlarge the number of suggestions. 
This kind of autocomplete and error-correction type-hinting system would be beneficial for various use-cases, e.g., in educational software or for search engines as a pre-processing step of the input.

\begin{figure}[t]
	\vspace*{-0.1cm}
	\centering
  	\includegraphics[width=0.423\textwidth]{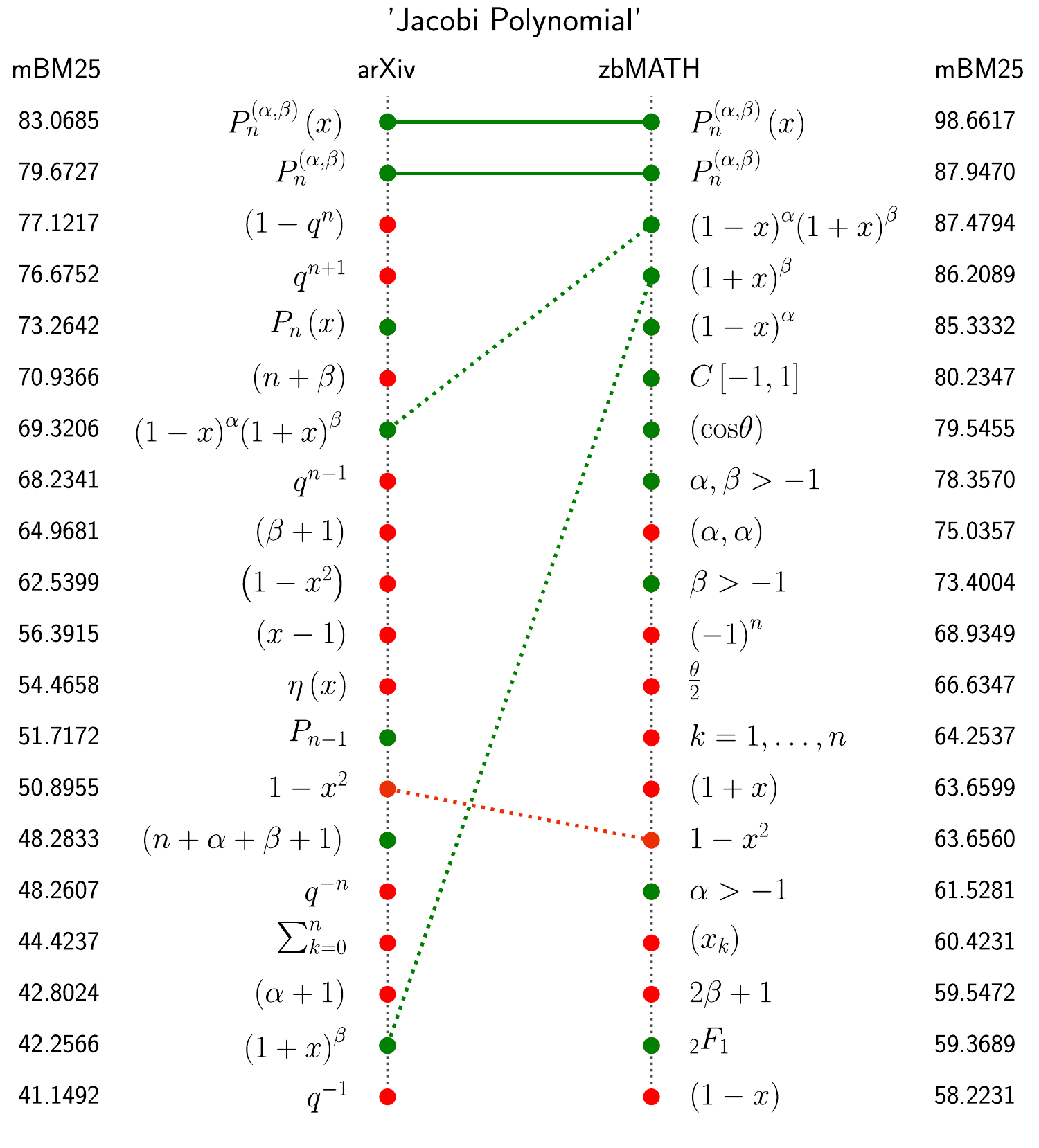}
  	\vspace*{-0.5cm}
	\caption{The top ranked expression for `\textit{Jacobi polynomial}' in arXiv and zbMATH. For arXiv, 30 documents were retrieved with a minimum hit frequency of 7.}
	\label{fig:tfidf-jacobi}
	\vspace*{-0.4cm}
\end{figure}

\noindent
\textbf{Plagiarism Detection Systems:} As previously mentioned, plagiarism detection systems~\cite{MeuschkeSHSG17,Meuschke2019,Schubotz2019} would benefit from a system capable of distinguishing conventional from uncommon notations. 
The approaches described by Meuschke et al.~\cite{Meuschke2019} outperform existing approaches by considering frequency distributions of single identifiers (expressions of complexity one). 
Considering that single identifiers make up only $0.03\%$ of all unique expressions in arXiv, we presume that better performance can be achieved by considering more complex expressions.
The conferred string representation also provides a simple format to embed complex expressions in existing learning algorithms.

Expressions with high complexities that are shared among multiple documents may provide further hints to investigate potential plagiarisms. 
For instance, the most complex expression that was shared among three documents in arXiv was Equation~\eqref{eq:hardy-littlewood}.
A complex expression being identical in multiple documents could indicate a higher likelihood of plagiarism.
Further investigation revealed that similar expressions, e.g., with infinite sums, are frequently used among a larger set of documents. 
Thus, the expression seems to be a part of a standard notation that is commonly shared, rather than a good candidate for plagiarism detection.
Resulting from manual investigations, we could identify the equation as part of a concept called \textit{generalized Hardy-Littlewood inequality} and Equation~\eqref{eq:hardy-littlewood} appears in the three documents~\cite{Pell1,Pell2,Pell3}. 
All three documents shared one author in common. 
Thus, this case also demonstrates a correlation between complex mathematical notations and authorship.

\noindent
\textbf{Semantic Taggers and Extraction Systems:} We previously mentioned that semantic extraction systems~\cite{Schubotz16,Kristianto2017,Schubotz2017} and semantic math taggers~\cite{ChienC15,POM-Tagger} have difficulties in extracting the essential components (MOIs) from complex expressions. 
Considering the definition of the Jacobi polynomial in Equation~\eqref{eq:jacobi-def}, it would be beneficial to extract the groups of tokens that belong together, such as $\StandardJacobi$ or $\Gamma(\alpha + m + 1)$. 
With our proposed search engine for retrieving MOIs, we are able to facilitate semantic extraction systems and semantic math taggers.
Imagine such a system being capable of identifying the term `Jacobi polynomial' from the textual context. 
Figure~\ref{fig:tfidf-jacobi} shows the top relevant hits for the search query `Jacobi polynomial' retrieved from zbMATH and arXiv. 
The results contain several relevant and related expressions, such as the constraints $\alpha, \beta > -1$ and the weight function for the Jacobi polynomial $(1-x)^{\alpha}(1+x)^{\beta}$, which are essential properties of this orthogonal polynomial. 
Based on these retrieved MOIs, the extraction systems can adjust its retrieved math elements to improve precision, and semantic taggers or a tokenizer could re-organize parse trees to more closely resemble expression trees.

\section{Conclusion \& Future Work}\label{sec.concloutl}
In this study we showed that analyzing the frequency distributions of mathematical expressions in large scientific datasets can provide useful insights for a variety of applications. 
We demonstrated the versatility of our results by implementing prototypes of a type-hinting system for math recommendations, an extension of zbMATH's search engine, and a mathematical retrieval system to search for topic-specific MOIs. 
Additionally, we discussed the potential impact and suitability in other applications, such as 
math search engines, plagiarism detection systems, and semantic extraction approaches. 
We are confident that this project lays a foundation for future research in the field of MathIR.

We plan on developing a web application which would provide easy access to our frequency distributions, the MOI search engine, and the type-hinting recommendation system. 
We hope that this will further expedite related future research projects.
Moreover, we will use this web application for an online evaluation of our MOI retrieval system. Since the level of agreement among annotators will be predictably low, an evaluation by a large community is desired.

In this first study, we preserved the core structure of the \MathML{} data which provided insightful 
information for the \MathML{} community. However, this makes it difficult to properly merge formulae.
In future studies, we will normalize the \MathML{} data via MathMLCan~\cite{MathMLCan}.
In addition to this normalization, we will include wildcards for investigating distributions of 
formula patterns rather than exact expressions. 
This will allow us to study connections between math objects, e.g., between $\Gamma(z)$ and $\Gamma(x+1)$. This would further improve our recommendation system and would allow for the identification of regions for parameters and variables in complex expressions.

\noindent\textbf{Acknowledgments} Discovering Mathematical Objects of Interest was supported by the German Research Foundation (DFG grant GI-1259-1).

\printbibliography%
\end{document}